\documentstyle[aaspp4,psfig,11pt]{article}

\received{2000 April 15}
\accepted{2000 July 11}
\slugcomment{Accepted by the {\it Astrophysical Journal} on July 11, 2000}

\begin{document}

\def\ISO{{\it ISO}}
\def\etal{ et al. }
\def\arcsec{\hbox{$^{\prime\prime}$}}
\def\simgt{\ga}
\def\simlt{\lower.5ex\hbox{$\; \buildrel < \over \sim \;$}}

\title{Infrared Spectroscopy of Molecular Supernova Remnants}

\author{William T. Reach  and 
          Jeonghee Rho }

\affil{Infrared Processing and Analysis Center, 
California Institute of Technology,
Pasadena, CA 91125}
\authoremail{reach@ipac.caltech.edu}

\begin{abstract}

\def\textabs{

We present Infrared Space Observatory spectroscopy of sites in the supernova
remnants W28, W44, and 3C391, where blast waves are impacting molecular clouds. 
Atomic fine-structure lines were detected from C, N, O, Si, P, and Fe. The S(3)
and S(9) lines of H2 were detected for all three remnants. The observations
require both shocks into gas with  moderate (~ 100 /cm3) and high (~10,000 /cm3)
pre-shock densities, with the moderate density shocks producing the ionic lines
and the high density shock producing the molecular lines. No single shock model
can account for all of the observed lines, even at the order of magnitude level.
We find that the principal coolants of radiative supernova  shocks in
moderate-density gas are the far-infrared continuum from dust grains surviving
the shock, followed by collisionally-excited [O I] 63.2 and [Si II] 34.8 micron
lines. The principal coolant of the high-density shocks is collisionally-excited
H2 rotational and ro-vibrational line emission. We systematically examine the
ground-state fine structure of all cosmically abundant elements, to explain the
presence or lack of all atomic fine lines in our spectra in terms of the atomic
structure, interstellar abundances, and a moderate-density, partially-ionized
plasma. The [P II] line at 60.6 microns is the first known astronomical
detection. There is one bright unidentified line in our spectra, at 74.26
microns. The presence of bright [Si II] and [Fe II] lines requires partial
destruction of the dust. The required gas-phase abundance of Fe suggests 15-30%
of the Fe-bearing grains were destroyed. The infrared continuum brightness
requires ~1 Msun of dust survives the shock, suggesting about 1/3 of the dust
mass was destroyed, in agreement with the depletion estimate and with theoretical
models for dust destruction.

}

We present {\it Infrared Space Observatory}
spectroscopy of sites in the supernova remnants W~28, W~44, and 3C~391,
where blast waves are impacting molecular clouds. 
The complete wavelength range from 42 to 188 $\mu$m was observed with the 
Long Wavelength Spectrometer,
as well as  narrow ranges centered on 4.695, 9.665, 25.98 and 34.82 $\mu$m
with the Short Wavelength Spectrometer.
Atomic fine-structure lines were detected from (in order of atomic number): 
C$^+$, N$^+$, N$^{++}$, O$^0$, O$^{++}$, O$^{+++}$,  
Si$^+$, P$^+$, and Fe$^+$.
The two lines of H$_2$ that we observed, S(3) and S(9) were detected for 
all three remnants.
The observations require both shocks into gas with 
moderate ($\sim 10^{2}$ cm$^{-3}$) and high ($\sim 10^{4}$ cm$^{-3}$) 
pre-shock densities,
with the moderate density shocks producing the ionic lines 
 and the high density shock producing the
molecular lines.
No single shock model can account for all of the observed lines, 
even at the order of magnitude level.
We find that the principal coolants of radiative supernova 
shocks in moderate-density gas are the far-infrared continuum from
dust grains surviving the shock, followed by collisionally-excited
[\ion{O}{1}] 63.2 $\mu$m and [\ion{Si}{2}] 34.8 $\mu$m lines.
The principal coolant of the high-density shocks is collisionally-excited
H$_2$ rotational and ro-vibrational line emission.
We systematically examine the ground-state fine structure of all
cosmically abundant elements, to explain
the presence or lack of all atomic fine lines in our spectra
in terms of the atomic structure, interstellar abundances, and
a moderate-density, partially-ionized plasma.
The [\ion{P}{2}] line at 60.6 $\mu$m is the first known astronomical detection,
but its brightness can be explained using the solar abundance of P.
There is only one, bright unidentified line in our spectra, at 74.26 $\mu$m;
as there is no plausible atomic fine-structure line at this wavelength, 
we suggest this line is molecular. 
The presence of bright [\ion{Si}{2}] and [\ion{Fe}{2}] lines requires
partial destruction of the dust. The required gas-phase abundance of
Fe suggests 15--30\% of the Fe-bearing grains were destroyed.
Adding the Si and Fe gas mass, and correcting for the mass of other elements
normally found in dust, we find $\sim 0.5 M_\odot$ of dust vapors 
from the shocked clump 3C~391:BML. The infrared continuum brightness requires $\sim 1 M_\odot$ of
dust survives the shock, suggesting about 1/3 of the dust mass was
destroyed, in agreement with the depletion estimate and with theoretical
models for dust destruction.


\end{abstract}
                            

\section{Introduction}

Infrared spectroscopy is a particularly useful tool for studying
supernova remnants (and other shocks) in dense gas
for the following reasons. (1) Infrared atomic fine-structure
lines are produced by gas with
the physical conditions expected behind radiative shocks. 
Dense, shocked gas rapidly cools to 
temperatures of $10^2$--$10^3$~K, where infrared transitions are excited.
This cool layer dominates the shocked column density, while 
the layer that emits optical to X-ray lines is extremely narrow.
The kinetic
temperature of the cooling post-shock gas is perfectly suited to
the energy levels of infrared fine-structure lines from atoms and
ions and infrared vibrational, ro-vibrational, and quadrupole
vibrational (H$_2$) lines from molecules. 
(2) Infrared fine-structure lines 
trace the ground-state populations of
the all of the abundant	elements that have electronic
ground-states with non-zero angular momentum. 
By measuring the brightness of transitions
directly to the ground-state or within a narrow multiplet of energy 
levels including the ground-state, a reasonably accurate measure of
the column density of each species can be determined. Using
pairs of lines from different elements due to energy levels with
similar spacings, it is possible to measure relative abundances with
confidence. 
(3) Far-infrared fine-structure lines are not affected
by extinction. The light produced by post-shock gas
is significantly affected by extinction from dust in the pre-shock 
cloud and the intervening medium along the line of sight. Extinction
effects are severe in the ultraviolet and also in the optical and
near-infrared for inner-galaxy supernova remnants, where the line-of-sight
column density often exceeds $10^{22}$ cm$^{-2}$, but it
is negligible in the far-infrared. 
Many supernova remnants interacting with clouds are known
only from their radio or hard X-ray emission, despite being copious
producers of soft X-rays and optical line emission, because they are
too distant or they are within or behind a molecular cloud.

Recent advances in infrared astronomy have opened the far-infrared
window to spectroscopy of faint emission. The observations used
in this paper were made with the Long-Wavelength
Spectrometer (LWS; \cite{clegg}) aboard the {\it Infrared Space
Observatory} (\ISO; \cite{kessler}), whose mission lasted from
1996--1997. A complete spectrum with 
the LWS, covering the 40--190 $\mu$m wavelength region with a resolution
of $0.29$--$0.6$ $\mu$m, took less than half an hour to reach a sensitivity
adequate to detect atomic fine-structure lines and molecular rotational
lines for a variety of astronomical sources. Instruments planned for
the Stratospheric Observatory for Infrared Astronomy (SOFIA; \cite{becklin};
expected to begin observing in 2002)
will open opportunities for far-infrared spectroscopic observations 
over most of the far-infrared wavelength range---specifically, the
part not affected by absorption and emission
by very abundant molecules in the residual atmosphere such as H$_2$O.
In the somewhat-more-distant future, instruments aboard the
Far-Infrared Space Telescope (FIRST; \cite{firstref}; anticipated launch
2007) will allow continuous spectral coverage to even deeper levels for
extended sources. 

The goal of this paper is to present our {\it ISO} LWS and SWS
spectral observations
of the molecular supernova remnants W~44, W~28, and 3C~391,
and to illustrate from basic physical principles
why the spectral lines we detected are bright.
Shocked gas in supernova remnants emits virtually all fine-structure lines
from the abundant elements.
We show the inferred elemental abundances after impact from the
supernova shock, and we discuss implication of  grain destruction
for elements such as Fe, Si, and P  which were locked in
grains.

This paper continues our investigation of infrared emission from
molecular supernova remnants.
The brightness of the [\ion{O}{1}] 63.2 $\mu$m line for W~44 and 3C~391
was presented in our
first paper (\cite{rr96}).
The far-infrared molecular emission of 
OH, CO, and H$_2$O was presented in the second paper (\cite{rr98}).
The millimeter-wave molecular emission of CO, CS, HCO$^+$ was presented 
in the third paper for 3C~391 (\cite{rr99}), and in this paper
we also present some molecular results for W~44 from new IRAM 30-m observations.
Two recent and closely-related studies of molecular emission were presented
by Seta et al. (1998) for W~44 and Arikawa et al. (1999)
for W~28. The
millimeter-wave observations show beyond doubt that all three of
the supernova remnants discussed in this paper are interacting
with molecular clouds.

\epsscale{0.5}
\plotone{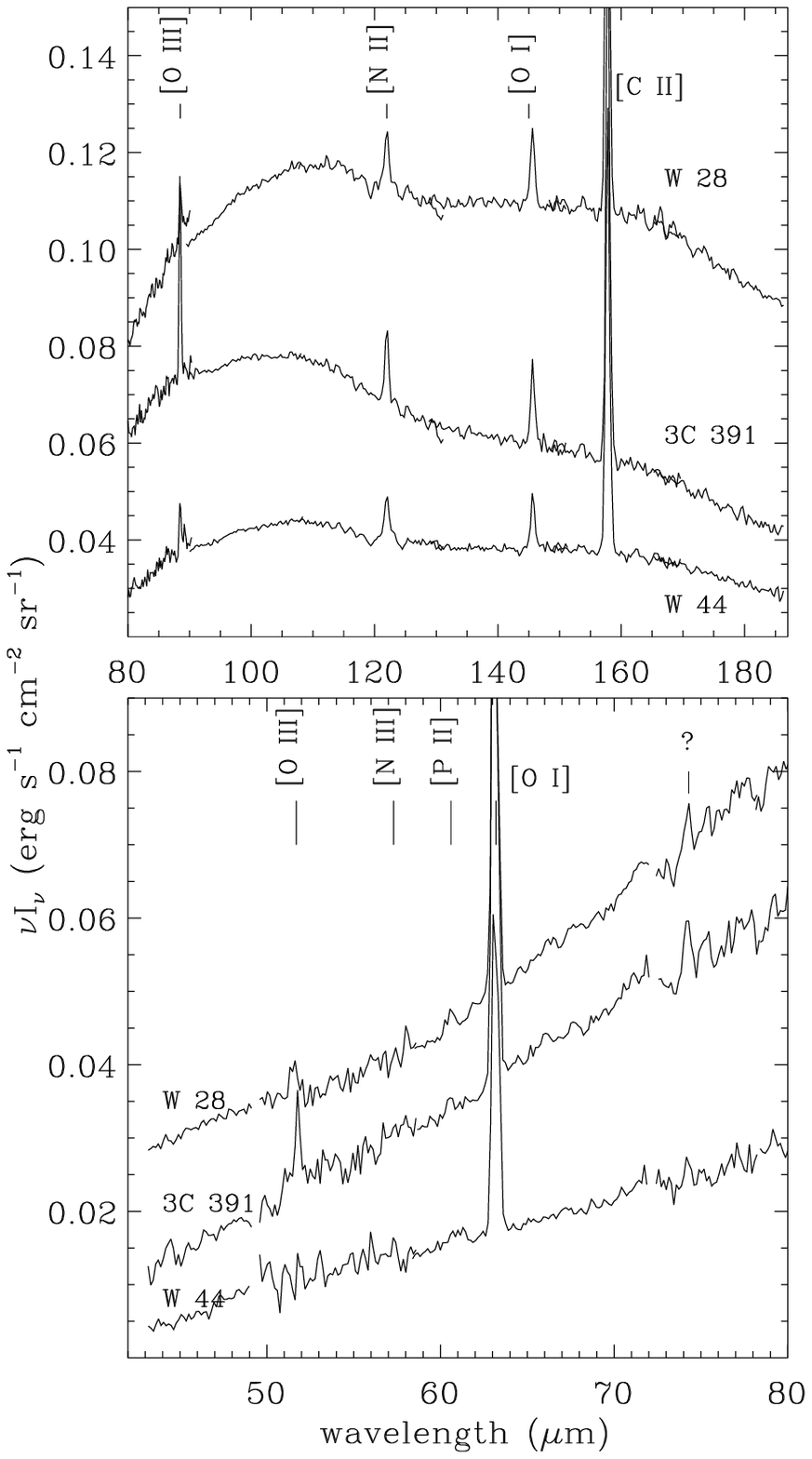}
\epsscale{1}
\figcaption{\it {\it ISO} LWS spectra of molecular shocks in the supernova
remnants W~28, 3C~391, and W~44. The lower and upper panels show the
wavelength ranges 42--80 and 80--187 $\mu$m, respectively. The ordinate 
in both panels is $\nu I_\nu$; to measure line fluxes note that the
resolution of the LWS is $\Delta\nu=0.29$ $\mu$m for the top panel and 0.6 $\mu$m
for the bottom panel. Note that no reference spectrum has been 
subtracted, so the continuum and [\ion{C}{2}] line 
are partly due to line-of-sight dust and gas.
The other bright atomic fine structure lines are due to the remnants.
Some of the minor wiggles at long wavelengths are molecular lines 
(Reach \& Rho 1998),
while detector noise is significant at the shortest wavelengths.
The unidentified line is labeled with a `?'.\label{allspec}}

\begin{table}
\caption[]{Observed far-infrared Atomic fine-structure lines}\label{spectab} 
\begin{flushleft} 
\begin{tabular}{llrrrr} 
\hline
 & & & \multicolumn{3}{c}{ Line Brightness ($10^{-4}$ erg~s$^{-1}$cm$^{-2}$sr$^{-1}$) } \\ \cline{4-6}
\multicolumn{1}{c}{ion} & \multicolumn{1}{c}{transition} & \multicolumn{1}{c}{wavelength} & 
	\multicolumn{1}{c}{W~28} & \multicolumn{1}{c}{W~44} & \multicolumn{1}{c}{3c~391:BML} 
	\\ \hline \hline
C$^+$ & $^2$P$_{\frac{3}{2}\rightarrow\frac{1}{2}}$ & 157.741 & 
		$7.77\pm 0.08$ & $4.94\pm 0.04$ & $6.54\pm 0.08$ \\
O$^0$ & $^3$P$_{0\rightarrow 1}$ & 145.525 & 
		$0.75\pm 0.04$ & $0.50\pm 0.03$ & $0.90\pm 0.02$ \\
N$^+$ & $^3$P$_{2\rightarrow 1}$  & 121.898 & 
		$0.68\pm 0.04$ & $0.42\pm 0.04$ & $0.89\pm 0.02$ \\
O$^{++}$ & $^3$P$_{1\rightarrow 0}$ &  88.356 & 
		$0.38\pm 0.06$ & $0.36\pm 0.02$ & $1.49\pm 0.04$ \\
O$^0$ & $^3$P$_{1-2}$ &  63.184&
		$9.48\pm 0.02$ & $7.73\pm 0.04$ & $12.96\pm 0.02$ \\
P$^+$ & $^3$P$_{1\rightarrow 0}$ &  60.640 & 
		$0.06\pm 0.03$ & $<.10$ & $0.095\pm 0.025$ \\
N$^{++}$ & $^2$P$_{\frac{3}{2}\rightarrow\frac{1}{2}}$ & 57.340 &
		$0.093\pm 0.033$ & $< .14$ & $0.25\pm 0.04$ \\
O$^{++}$ & $^3$P$_{2\rightarrow 1}$ &  51.815 & 
		$0.23\pm 0.06$ & $<0.23$ & $0.75\pm 0.08$ \\
\hline
\end{tabular} 
\end{flushleft} 
\end{table}  

\raggedright

\section{Observations}

\subsection{Far-infrared spectral lines\label{sec:firobs}}

We obtained complete {\it ISO} LWS spectra of 1 position in each of three
supernova remnants as part of our program of observing
supernova-molecular cloud interactions (Reach \& Rho 1996, 1998, 1999).
The positions observed were the brightest OH masers (\cite{Frail96})
in W~28 ($18^h01^m52.3^s$ $-23^\circ19^\prime25^{\prime\prime}$) and
W~44 ($18^h56^m28.4^s$ $+1^\circ29^\prime59^{\prime\prime}$)
and the molecular peak in 3C~391:BML 
(18$^h$49$^m$21.9$^s$ -0$^\circ$57$^\prime$22$^{\prime\prime}$);
all coordinates are J2000. 
Figure~\ref{allspec} shows the {\it ISO} LWS spectrum from 42--188 $\mu$m
for all three remnants.
For each of the 10 LWS detectors, we removed
the fringes using the {\it ISO} Spectral Analysis 
Package\footnote{The ISO Spectral Analysis Package (ISAP) is a joint development by
the LWS and SWS Instrument Teams and Data Centers. Contributing institutes are
CESR, IAS, IPAC, MPE, RAL and SRON.}
and subtracted a constant (due to residual dark current). 
Table~\ref{spectab} shows a list of the identified atomic fine structure 
lines and their measured brightness.
In order of atomic number, we detected atomic fine-structure
lines from 
C$^+$, N$^+$, N$^{++}$, O$^0$, O$^{++}$, O$^{+++}$, and
P$^+$ in the LWS spectra.

Part of the bright continuum and [\ion{C}{2}] 157.7 $\mu$m line in Fig.~\ref{allspec} 
are due to dust and gas along the line of sight,
unrelated to the supernova remnant. 
At least part of the continuum is likely to be due to
dust grains surviving the shocks. For example, for W~44 we estimated 30\% (\cite{rr96})
based on comparison to a reference position; a spectral analysis of the continuum
is given in \S\ref{section:continuum} of this paper.
To estimate the line-of-sight contamination of the spectral line brightnesses,
we calculated the galactic line-to-continuum ratio using
the {\it COBE} Far-Infrared Absolute Spectrophotometer spectrum of the 
galactic plane at longitude $30^\circ$ (\cite{reachfiras}).
At the {\it ISO} LWS resolution, the expected line-to-continuum ratio due
to unrelated gas 
for [\ion{C}{2}] 157.7 $\mu$m is 2.2, while that for [\ion{N}{2}] at 121.9 $\mu$m
is 0.05. Comparing to our spectra, the 157.7 $\mu$m line
to continuum ratios are 2.4, 1.2, and 1.9, and the 121.9 $\mu$m line to 
continuum ratios are 0.19, 0.09, and 0.16, 
for W~44, W~28, and 3C~391, respectively.
Thus much of
the [\ion{C}{2}] $\mu$m line emission could be due to line-of-sight gas, 
while only a small fraction
the [\ion{N}{2}] (and essentially none of the other lines) are due to
line-of-sight gas. The subject of [\ion{C}{2}] emission being related to the shock
is addressed again in \S\ref{section:abundance}, focusing on the abundance of C.
An improved separation of the source and background brightness might be possible
by analyzing the spatial variation of the brightness, which is beyond the
scope of this paper.

\epsscale{0.5}
\plotone{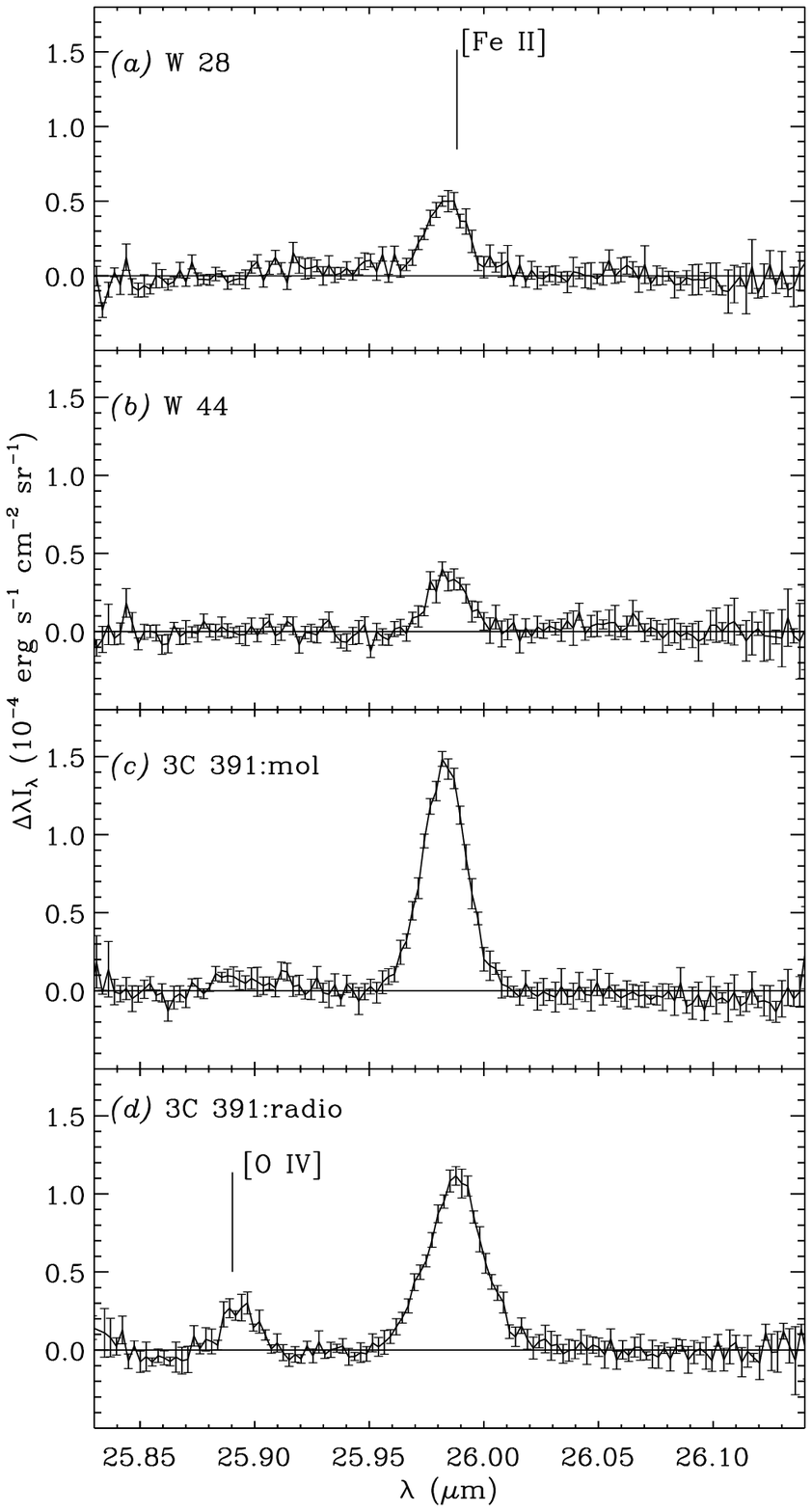}
\epsscale{1}
\figcaption{\it {\it ISO} SWS spectra of the ground-state [\ion{Fe}{2}] fine-structure
line for lines of sight toward molecular shocks in 
{\it (a)} W~28,
{\it (b)} W~44, and
{\it (c)} 3C~391; also the spectrum for a line of sight toward
the radio peak in 3C~391.
The molecular shock positions all show bright [\ion{Fe}{2}] lines.
The radio peak position in 3C~391 also shows a bright ground-state
fune structure line of [\ion{O}{4}]. 
The solid curves are Gaussian fits to the line with width fixed to the SWS grating
spectral resolution of $\Delta\lambda=0.018$ $\mu$m.
\label{swsfig1}}

\epsscale{0.5}
\plotone{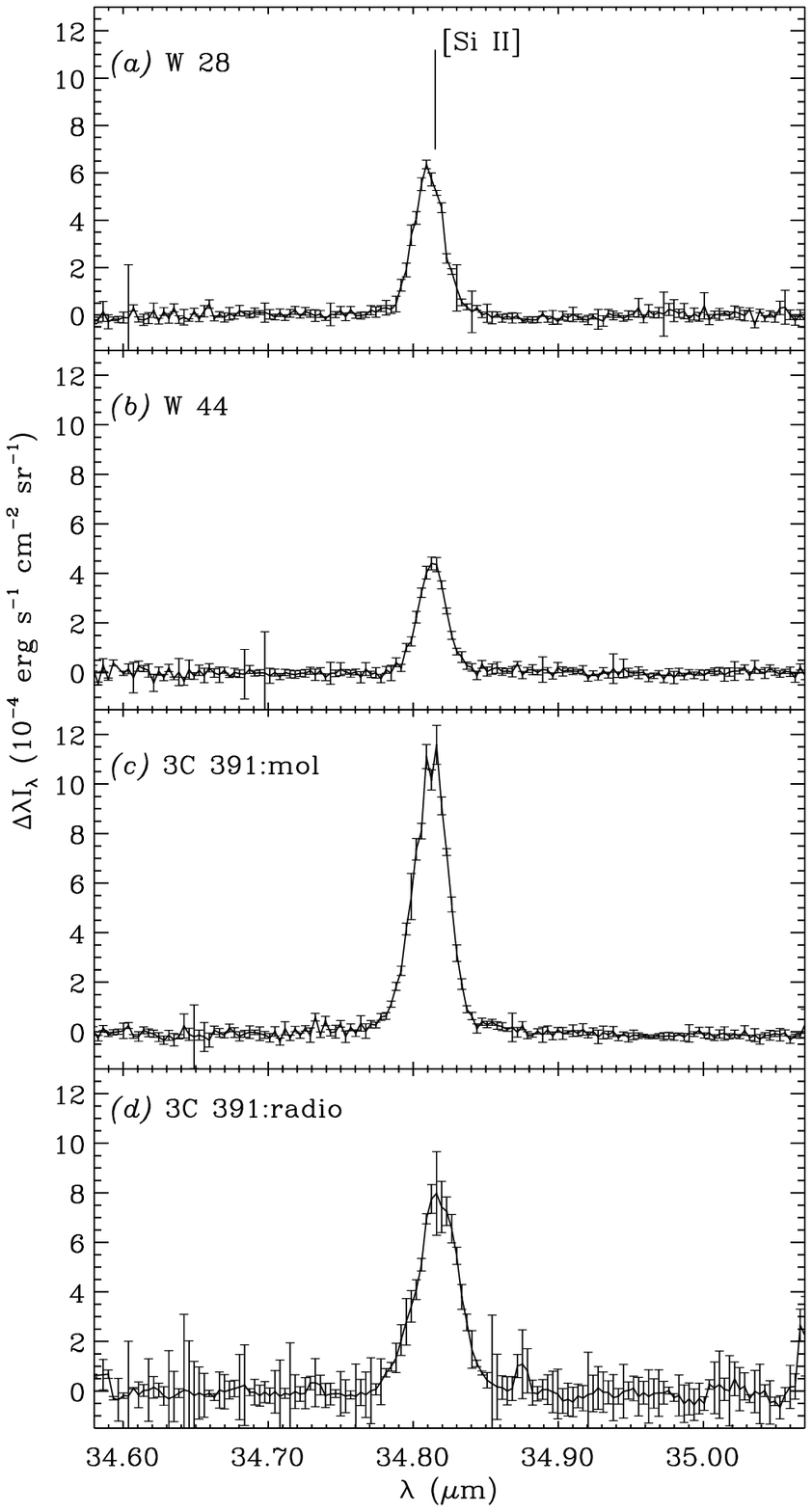}
\epsscale{1}
\figcaption{\it {\it ISO} SWS spectra of the ground-state [\ion{Si}{2}] fine-structure
line for lines of sight toward molecular shocks in 
{\it (a)} W~28,
{\it (b)} W~44, and
{\it (c)} 3C~391; also the spectrum for a line of sight toward
the radio peak in 3C~391.
The solid curves are Gaussian fits to the line with width fixed to the SWS grating
spectral resolution of $\Delta\lambda=0.022$ $\mu$m.
\label{swsfig2}}

\epsscale{0.5}
\plotone{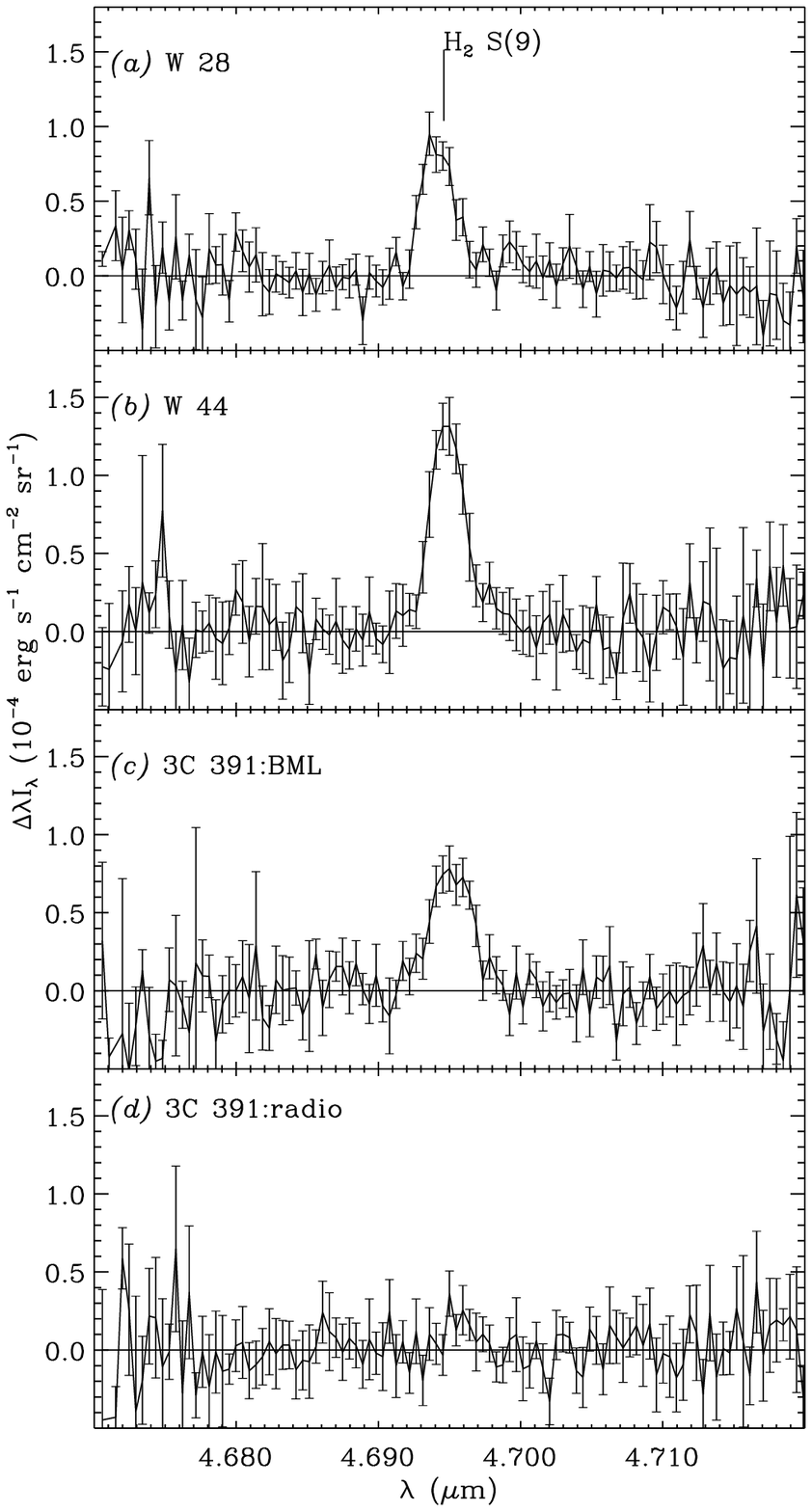}
\epsscale{1}
\figcaption{\it {\it ISO} SWS spectra of the H$_2$ S(9) rotational line 
toward molecular shocks in
{\it (a)} W~28,
{\it (b)} W~44, and
{\it (c)} 3C~391; also the spectrum for a line of sight toward
the radio peak in 3C~391.
The solid curves are Gaussian fits to the line with width fixed to the SWS grating
spectral resolution of $\Delta\lambda=0.0021$ $\mu$m.
\label{h2fig1}}

\epsscale{0.5}
\plotone{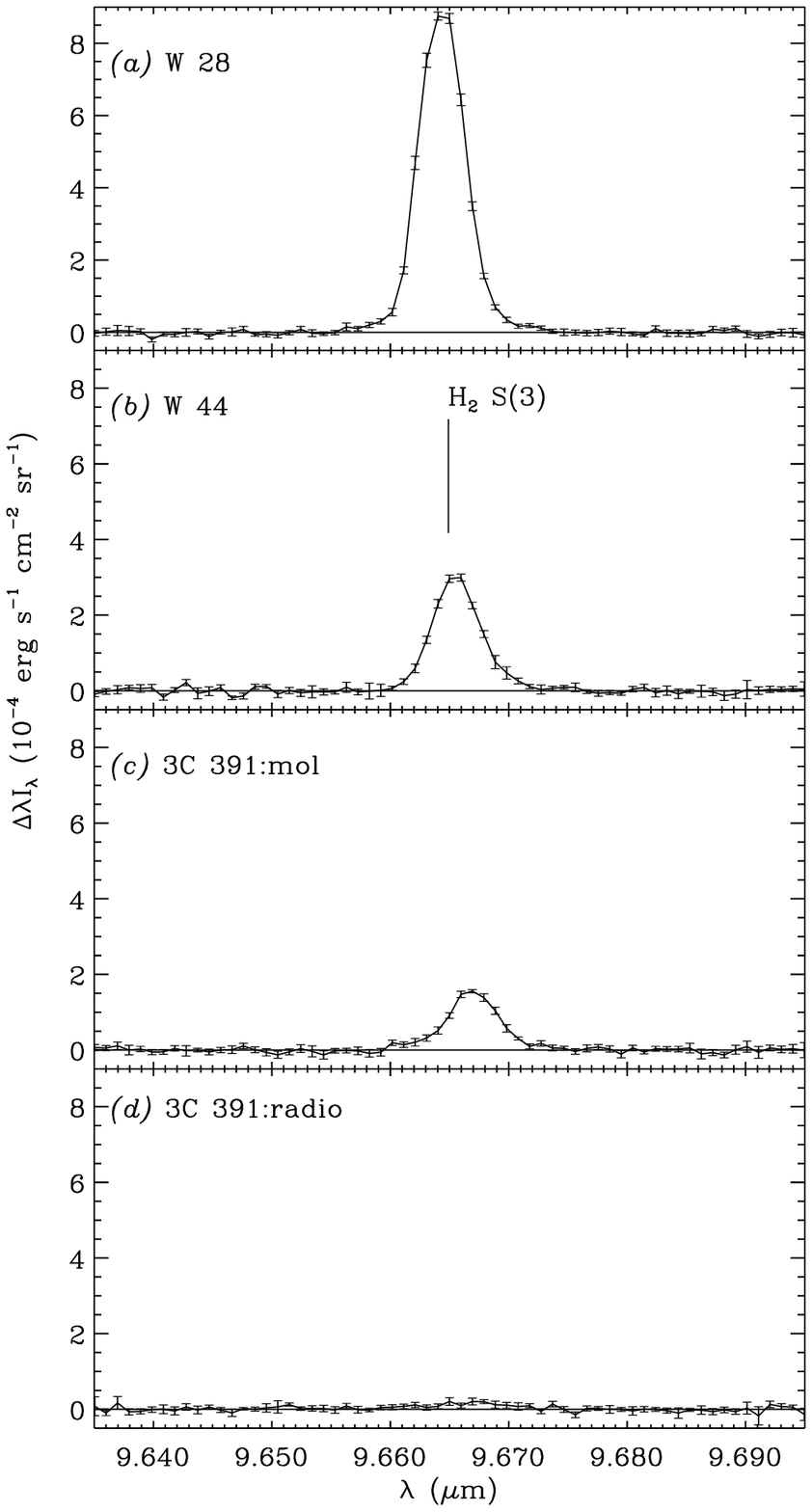}
\epsscale{1}
\figcaption{\it {\it ISO} SWS spectra of the H$_2$ S(3) rotational line 
toward molecular shocks in
{\it (a)} W~28,
{\it (b)} W~44, and
{\it (c)} 3C~391; also the spectrum for a line of sight toward
the radio peak in 3C~391.
The solid curves are Gaussian fits to the line with width fixed to the SWS grating
spectral resolution of $\Delta\lambda=0.0042$ $\mu$m.
\label{h2fig2}}

\begin{table}
\caption[]{SWS Spectral line results}\label{swstab} 
\begin{flushleft} 
\begin{tabular}{llrrrrr} 
\hline
 & & & \multicolumn{4}{c}{ Line Brightness ($10^{-4}$ erg~s$^{-1}$cm$^{-2}$sr$^{-1}$) } \\ \cline{4-7}
\multicolumn{1}{c}{ion} & \multicolumn{1}{c}{transition} & \multicolumn{1}{c}{wavelength} & 
	\multicolumn{1}{c}{W~28} & \multicolumn{1}{c}{W~44} & \multicolumn{1}{c}{3c~391:BML} 
	& \multicolumn{1}{c}{3C~391:radio} 
	\\ \hline \hline
Fe$^+$ & $^6$D$_{\frac{7}{2}\rightarrow\frac{9}{2}}$ & 25.988 &
		$0.66\pm 0.05$ & $0.41\pm 0.04$ & $1.9\pm 0.1$ & $1.8\pm 0.1$ \\
O$^{+++}$ & $^2$P$_{\frac{3}{2}\rightarrow\frac{1}{2}}$ & 25.890 &
		$<0.08$ & $<0.08$ & $<0.08$ & $0.12$: \\
Si$^+$ & $^2$P$_{\frac{3}{2}\rightarrow\frac{1}{2}}$ & 34.815 &
		$7.3\pm 0.1$ & $4.8\pm 0.1$ & $15\pm 0.2$ & $13\pm 0.2$ \\
H$_2$ & S(3) & 9.665 &
		$10.2\pm 0.1$ & $3.6\pm 0.1$ & $1.95\pm 0.05$ & $0.32\pm 0.06$ \\
H$_2$ & S(9) & 4.695 &
		$1.2\pm 0.1$ & $1.9\pm 0.1$ & $1.3\pm 0.1$ & $<0.2$ \\

\hline
\end{tabular} 
\end{flushleft} 
\end{table}  

\raggedright

\subsection{Mid-infrared Si, Fe, and H$_2$ lines\label{section:h2}}

We used the {\it ISO} SWS (\cite{swsref}) to observe narrow ranges centered 25.98 and 
34.82 $\mu$m for the ground-state transitions of [\ion{Fe}{2}] and [\ion{Si}{2}].
The spectra are shown in Figures~\ref{swsfig1} and~\ref{swsfig2},
respectively.
At the same time, we obtained spectra of the H$_2$ S(9) and S(3) lines;
the spectra of these lines are shown in Figures~\ref{h2fig1} and~\ref{h2fig2},
respectively.
The brightnesses of the lines are summarized in Table~\ref{swstab}.
The SWS observations were toward the three lines of sight listed above,
plus one extra position---toward the bright, northwestern radio ridge and
a bright peak in the [\ion{O}{1}] 63 $\mu$m strip map (position {\it b} in
Fig. 3 of \cite{rr96}):
3C~391:radio (18$^h$49$^m$19.4$^s$ -0$^\circ$55$^\prime$05$^{\prime\prime}$).
The two positions in 3C~391 have similar spectra of ionic fine structure lines:
the ratio of [\ion{Fe}{2}]/[\ion{Si}{2}] is nearly identical, 
although the [\ion{O}{4}] is relatively brighter toward 3C~391:radio by at least
a factor of 3. In contrast, the brightness of the H$_2$ lines, relative to the ionic 
lines, is more than a factor of 10 lower for 3C~391:radio than for
3C~391:BML. 
This is consistent with the millimeter-wave molecular lines revealing
no evidence for shocked molecular gas toward 3C~391:radio, 
while bright and broad shocked molecular lines were detected 
toward 3C~391:BML (Figs. 2 and 4 of \cite{rr99}).
The spectra of these two positions serve a valuable
comparison and diagnostic tool, because one of them is dominated by ionic lines, while the
other has significant molecular emission.

\epsscale{0.8}
\plotone{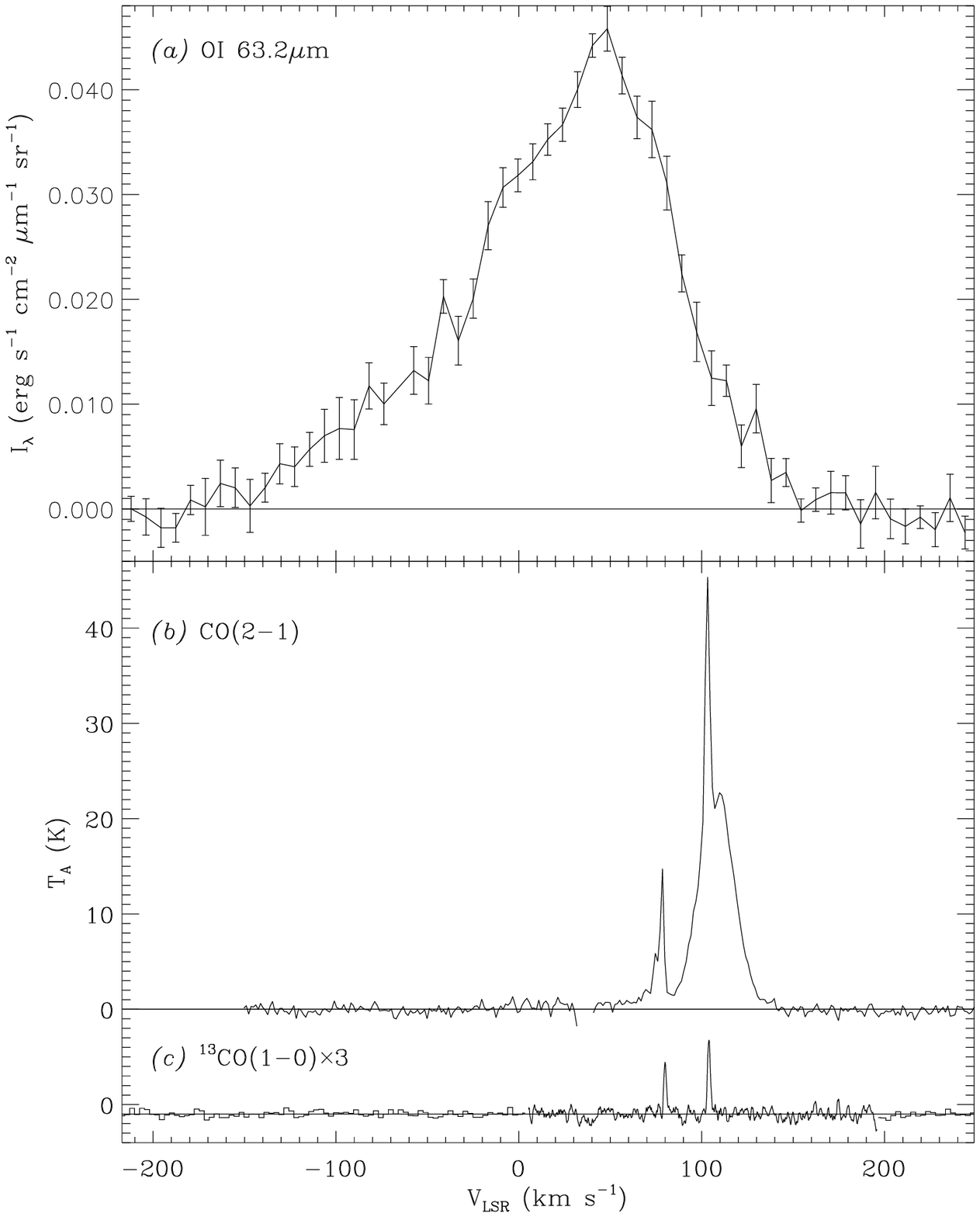}
\epsscale{1}
\figcaption{\it {\it (a)} High-resolution {\it ISO} LWS spectrum of the \ion{O}{1} line
toward 3C~391:BML. The instrumental resolution is 36 km~s$^{-1}$, so the
line is clearly resolved. {\it (b)} and {\it (c)} Spectra of the 
CO($2\rightarrow 1$) and $^{13}$CO($1\rightarrow 0$) 
lines for the same line of sight, obtained with the IRAM 30-m telescope.
The $^{13}$CO line arises only from unshocked gas, while the 
CO($2\rightarrow 1$) line has a wide component from the shocked
gas as well as narrow components from the unshocked gas.
\label{fp3c391}}

\epsscale{0.8}
\plotone{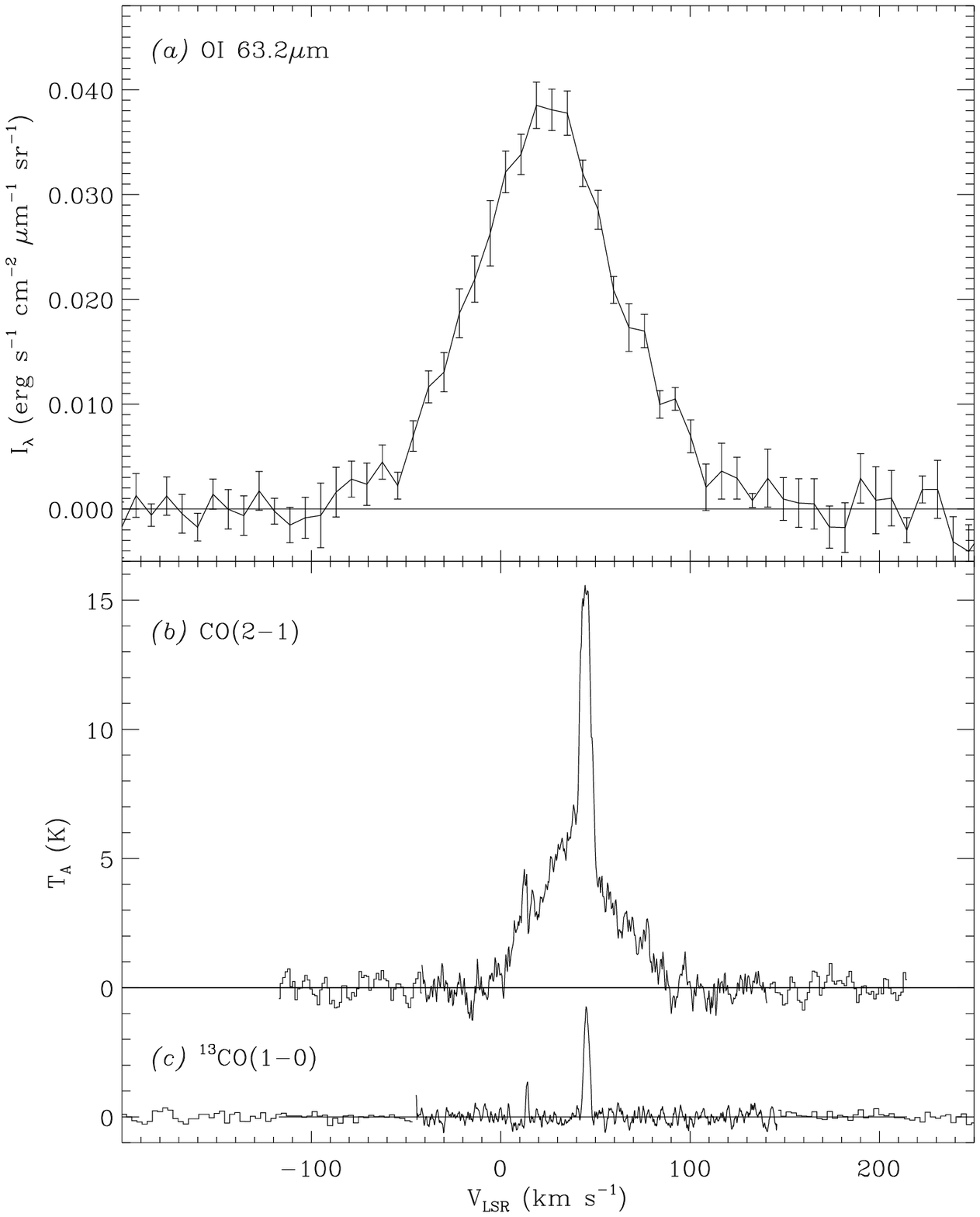}
\epsscale{1}
\figcaption{\it {\it (a)} High-resolution {\it ISO} LWS spectrum of the \ion{O}{1} line
toward W44. {\it (b)} and {\it (c)} Spectra of the 
CO($2\rightarrow 1$) and $^{13}$CO($1\rightarrow 0$) 
line for the same line of sight, obtained with the IRAM 30-m
telescope.
\label{fpw44}}

\raggedright

\subsection{High-resolution [\ion{O}{1}] and CO spectra\label{fabry}}

We used the Fabry-Perot in the LWS to obtain high-resolution spectral
profiles of the [\ion{O}{1}] 63 $\mu$m lines for 3C~391:BML and
W~44. The positions are the same as those where the complete LWS spectra
and the SWS line spectra were taken, and they correspond to peaks in the
[\ion{O}{1}] strip maps presented in paper 1\footnote{
Please note that the apparent 400 km~s$^{-1}$ shift in line center 
in 3C~391 from one side of
the remnant to the other, mentioned in paper 1, is apparently an instrumental artifact
due to mapping in the dispersion direction of the grating. The observations
presented in this paper, with the Fabry-Perot, are considered reliable measurements
of the line profile.}.
The instrumental resolution of the LWS Fabry-Perot is 36 km~s$^{-1}$.
The two spectra are shown in Figures~\ref{fp3c391} and~\ref{fpw44}.
The lines are clearly resolved, with a full width at half maximum 
(FWHM) of 100 km~s$^{-1}$. 
The lower panels of Figs.~\ref{fp3c391} and~\ref{fpw44} show the 
CO($2\rightarrow 1$) and $^{13}$CO($1\rightarrow 0$) line profiles for 
the same lines of sight
as the \ion{O}{1} spectra,
obtained with the IRAM 30-m telescope (\cite{wild}). The observing techniques are
described in our previous paper, where the 3C~391 observations were
presented (\cite{rr99}); the W~44 observations are new.
The line profiles are all clearly, very different.
The $^{13}$CO lines have a narrow (1.5 km~s$^{-1}$ FWHM) velocity dispersion, 
and they trace the column density, so they arises from large clouds of
cold, unshocked or unrelated gas.
Although the CO($2\rightarrow 1$) lines also have narrow components, identical to
the $^{13}$CO lines, from the unshocked or unrelated gas,
the CO($2\rightarrow 1$) lines are dominated by a broad component
(20--50 km~s$^{-1}$ FWHM).
The broad component was seen in the CO, CS, and HCO$^+$ observations of 3C~391,
with relatively more emission from the higher-excitation and higher dipole
moment lines (\cite{rr99}). Similarly, the CO observations of W~28 showed
that the broad component was much brighter than the narrow component in
the $3\rightarrow 2$ line as compared to the $1\rightarrow 0$ line
(\cite{arikawa}).
The broad components of the molecular lines are still not as broad as
the [\ion{O}{1}] line, suggesting that each of the spectral lines in
Figs.~\ref{fp3c391} and~\ref{fpw44} trace different regions.

Inspecting the [\ion{O}{1}] line profiles, there is only slight evidence for
structure. The lines from both remnants can be reasonably represented by 
gaussians with FWHM $\sim 100$ km~s$^{-1}$. 
For 3C~391:BML, the line profile is somewhat better
represented by multiple components: an unresolved component centered near 
50 km~s$^{-1}$ and a broad, non-Gaussian component centered near 0 km~s$^{-1}$
with 120--150 km~s$^{-1}$. The integrated line brightness is dominated by the 
wider component. The unresolved component contributes only about 10\% of the
line integral, and it may be associated with preshock or unrelated gas,
which would agree with the brightness of the `off-remnant' positions from the 
[\ion{O}{1}] strip map (\cite{rr96}). 
Neither 3C~391 nor W~44 show evidence for
self-absorption in the line profiles. This was a particular concern
in interpreting the line brightnesses for molecular supernova remnants,
both because of the possibility of dense pre-shock clouds and
the large column densities of cold gas along the lines of
sight to the remnants. 
A single cold cloud might be expected to yield an essentially
unresolved (Gaussian) absorption dip, which does not occur. 
The foreground gas would be expected to absorb in approximate
proportion to the line of sight column density. After inspecting
the \ion{H}{1} 21-cm profile for these lines of sight, we found no
correlation that could be due to absorption dips. 
Instead, it is likely that the
line-of-sight gas contributes slightly to the {\it emission}. The level
of [\ion{O}{1}] emission from the line-of-sight gas is most easily
estimated from the reference positions just outside the remnant,
where it was found to be quite faint (\cite{rr96}). Therefore,
we interpret the [\ion{O}{1}] line profiles in Figs.~\ref{fp3c391}
and~\ref{fpw44} as emission from the shocked gas. This inference
yields a direct measure of the shock velocity of 100--150 km~s$^{-1}$, 
albeit in projection along the line of sight. 
There are likely to be multiple shocks within
the beam, and the FWHM of the line profile is approximately equal to
the `typical' shock velocity.

Comparing the [\ion{O}{1}] and CO line profiles, we can partially
describe the three-dimensional geometry of the shocks. For 3C~391,
note that the centroids 
of the [\ion{O}{1}] and CO lines are very different. The CO line
occurs at nearly the systemic velocity at the remnant's distance,
with the broad component due to the shocked gas very slightly blueshifted 
from the narrow line from the preshock gas. 
The [\ion{O}{1}] line, on the other hand, is significantly red-shifted 
from the pre-shock gas. Thus the shock that produces the
[\ion{O}{1}] line has a significant velocity component toward us;
that is, the shock is propagating partially along the line of sight.
This suggests that the 3C~391 progenitor exploded on the far side
of its parent molecular cloud, driving the dense blast wave 
partially in our direction.
For W~44, the centroids of the [\ion{O}{1}] and CO lines are 
similar, suggesting that the shocks are either perpendicular to the
line of sight, or that the lines come from similar geometries.
One straightforward conclusion from comparing the line profiles, both
in centroid and width, is
that the [\ion{O}{1}] line does not arise exclusively
from slow shocks into very dense gas such as produces the CS, OH,
and H$_2$O emission (\cite{rr99}).
Instead, most of the [\ion{O}{1}] emission must arise from faster shocks 
into moderate-density gas.


\section{Comparison of detected lines to the periodic table}

\subsection{Atomic fine-structure lines: Basic principles}

The three basic principles that determine which spectral lines will be
bright in the spectrum of a parcel of gas are 
abundance, ionization, and excitation.
The abundances of the elements are given by Anders \& Grevesse (1989)
for the Solar System. In the interstellar medium, since part of the abundance
will be locked in solids, the gas phase abundance
will be smaller than Solar System abundance for some elements (\cite{sembachsavage}).
It is clear that we need only consider the top three rows of the
Periodic Table and the middle of the fourth row; all other elements
have negligible abundance. The ionization state of each element depends
on where the gas is located. We can consider three typical locations: the diffuse
interstellar medium (where all electrons bound by less than 13.6 eV
are removed), dark clouds (where elements are neutral), and hot
plasmas (where the higher ionization states are present). The excitation
of the atomic levels will depend on the density and temperature of the
region, but a `zero-th' order effect determines whether there are
any atomic fine structure lines in the ground electronic state. 
Elements and ions with zero angular momentum ($L=0$)
in their ground state electron configuration 
will not have the fine-structure lines considered here. 
For example, in spectroscopic notation,
S configurations have no lines, while doublet $^2$P configurations
have 1 line and triplet $^3$P configurations have 2 lines.
The fine-structure lines are magnetic dipole transitions, so they are
technically `forbidden lines', but with spontaneous radiative decay
rates of order $10^{-5}$ s$^{-1}$ and collisional de-excitation rates
of order $10^{-7}$ cm$^{-3}$~s$^{-1}$, they are easily excited under
astronomical conditions and are thermalized at densities 
$\simgt 10^{2}$ cm$^{-3}$ (for far-infrared lines) to 
$10^{4}$ cm$^{-3}$ (some mid-infrared lines).

\begin{deluxetable}{llllllll}
\tablecolumns{8}
\footnotesize
\tablecaption{`Periodic Table' for Astronomical Fine Structure Lines: $Z$=6--18 \tablenotemark{a} \label{tab:periodic}}
\tablewidth{0pt}
\tablehead{
\colhead{Element} & \colhead{Abundance\tablenotemark{b}}    & \colhead{$^2P_{1/2}$} & \colhead{$^3P_{0}$}         & \colhead{$^4S_{3/2}$} & \colhead{$^3P_2$}           & \colhead{$^2P_{3/2}$} & \colhead{$^1S_{3/2}$} }
\startdata
\hline
C  & -3.4                                & {\bf II}$_{\it 157.7}$    & I$_{610}^{370}$            &                       &                            &                       & \nl \hline
N  & -3.9                                &     III$_{\it 57.3}$     & { II}$_{205.9}^{\it 121.9}$ & {\bf I}...                 &                            &                       & \nl \hline
O  & -3.1                                &     IV$_{\it 25.9}$     &     III$_{\it 88.4}^{\it 51.8}$    & II...                & {\bf I}$_{\it 63.2}^{\it 145.5}$  &                       & \nl \hline
F  & -7.5                                &                        &                             &                       &        I$_{29.3}^{67.2}$  & {\bf I}$_{24.8}$     & \nl \hline
Ne & -3.9                                &                        &                             & {\it IV} ...          & {\it III}$_{ 15.6}^{36.0}$ &     II$_{ 12.8}$     & {\bf I}...\nl \hline
\tableline
Al & -5.5                                & I$_{89.2}$            &                             &                       &                            &                       & \nl \hline
Si & -4.4                                & {\bf II}$_{\it 34.8}$     & I$_{129.7}^{68.4}$         &                       &                            &                       & \nl \hline
P  & -6.4                                &     III$_{17.9}$     & {\bf II}$_{\it 60.6}^{32.9}$   & I...                 &                            &                       & \nl \hline
S  & -4.7                                &      IV$_{10.5}$     &     III$_{33.5}^{18.7}$   & {\bf II}...          & I$_{25.2}^{56.3}$         &                       & \nl \hline
Cl & -6.7                                &                        &                             &     III...          & {\bf II}$_{ 14.4}^{33.3}$  & I$_{11.3}$           & \nl \hline
Ar & -5.4                                &                        &                             &      IV...          &     III$_{8.99}^{21.8}$  &   II$_{\it 6.99}$        & {\bf I} ...
\tablenotetext{a}{
The headings of columns 3-8 give the spectroscopic notation $^xY_z$ for the ground state.
Roman numerals represent the ionization state (I=neutral, II=singly ionized, ...), with the dominant ionization 
state in the diffuse interstellar medium in {\it bold}.
Small numbers indicate the wavelength of the ground-state fine structure lines, in $\mu$m.
The transition to the true ground state is at the bottom and increasing energy levels are above,
when applicable.
All wavelengths of spectral lines that we detected are shown in {\it italics}.
Zero-spin ground states, which have no fine structure lines,
have `...' in their wavelength column.
}
\tablenotetext{b}{$A$ is the base-10 logarithm of the abundance relative to H.}
\enddata
\end{deluxetable}

\begin{deluxetable}{llllllll}
\tablecolumns{8}
\footnotesize
\tablecaption{`Periodic Table' for astronomical Fine Structure Lines: $Z$=24--28 \tablenotemark{a} \label{tab:periodictwo}}
\tablewidth{0pt}
\tablehead{
\colhead{Element} & \colhead{Abundance} & \colhead{23}     & \colhead{24} & \colhead{25} & \colhead{26} & \colhead{27} & \colhead{28}  }
\startdata
Cr & 6.3    & \def\skipme{{\it III $^5$D$_0$ \parbox{0.5cm}{$^{45.6}_{57.6}$  $^{82.9}_{161.9}$}} &}
              {\bf II$^6$S$_{5/2}$}...  & I$^7$S$_3$...                                                        &                                                                           &                                                                &                                  & \nl \hline
Mn & 6.5    &  III$^6$S$_{5/2}$... & {\bf II$^7$S$_3$}...                                                 & I $^6$S$_{5/2}$                                                           &                                                                &                                  & \vspace{3pt} \nl \hline
Fe & 4.5    & IV$^6$S$_{5/2}$...  & III$^5$D$_4$\parbox{0.5cm}{$^{105.4}_{51.7}$ $^{33.0}_{22.9}$} & {\bf II$^6$D$_{9/2}$} \parbox{0.5cm}{$^{87.3}_{\it 51.3}$ $^{35.3}_{\it 25.99}$} & I $^5$D$_4$\parbox{0.5cm}{$^{111.2}_{54.3}$ $^{34.7}_{24.0}$} &                                  & \vspace{3pt} \nl  \hline
Co & 7.1    &                             &                                                                        & III$^4$F$_{9/2}$ \parbox{0.5cm}{$^{24.1}_{16.4}$ $^{11.9}$}        & {\bf II$^3$F$_4$} $^{15.5}_{10.5}$                            & I$^4$F$_{9/2}$$^{16.9}_{12.3}$ & \nl  \hline
Ni & 5.7    &                             &                                                                        &                                                                           & III$^3$F$_4$ $^{11.0}_{7.34}$                           & {\bf II$^2$D$_{5/2}$}$_{6.64}$ & I$^3$F$_4$$^{11.3}_{7.51}$ 
\tablenotetext{a}{Column heading is the number of electrons. See caption for Tab.~\ref{tab:periodic}.} 
\enddata
\end{deluxetable}

In order to understand and predict which fine-structure lines are present in astronomical
spectra, we constructed a `periodic table for astronomical fine structure lines'.
Table~\ref{tab:periodic} contains the electron configurations and wavelengths for
the p-shell ions of the second and third rows of the Periodic Table
(C, N, O, Fe, Ne, Al, Si, P, S, Cl, and Ar), while
Table~\ref{tab:periodictwo} has the same information for the fourth row of the 
Periodic Table (Cr, Mn, Fe, CO, and Ni).
In both tables, the dominant ionization state in the diffuse interstellar
medium is shown in {\bf bold}. Spectra of regions behind shock fronts or with very
high radiation fields will contain the dominant ionization state as well as
and the ions to the left of the dominant ionization state. Spectra of
regions that are shielded from the interstellar radiation field will contain 
neutral forms and ions to the right of the dominant ionization state.
Some elements with negligible astronomical abundance are not listed; also, to reduce clutter,
unlikely ionization states of rare elements are not listed.
Alkaline metals (in low ionization states) have no $p$-shell electrons to make infrared 
fine-structure lines, so they are not listed.
These tables can be used both to predict which lines will be present for a given
gas and to decide which lines {\it should} be present when another line of the
same ion is detected. In all cases, the transitions are simple cascades down a
ladder, with the ground state at the bottom; therefore, the upper energy level
for a line can be calculated by summing the photon energies from a given line to
the bottom.

\subsection{Comparison of detected lines to the Periodic Table}

In the `periodic table for atomic fine-structure lines' (Tables~\ref{tab:periodic} 
and~\ref{tab:periodictwo}), the lines we detected from 
molecular supernova remnants observations are in {\it italics}. 
It appears that we detect all of the
available ground-state fine-structure lines of the astronomically
abundant molecules; 
the undetected ground-state lines are generally outside our wavelength range.
All abundant elements, C, N, O and Si, are detected as shown in
Tables~\ref{spectab} and~\ref{swstab}. 
Non-detections of F and Co lines are due to low abundances. 
Lines from high-ionization states were also not detected.
The only element with high abundance that we did not detect
is Ne, because our observations did not cover the wavelengths
of its ground-state fine-structure lines.
Our own observations contain the wavelengths of 12 fine-structure
transitions among energy levels within 600~K of the ground state, from
elements with abundance $> 10^{-7}$ relative to H in their expected
ionization states. We detect essentially all of these lines, as well as
lines from higher ionization states of the abundant elements: 
\ion{O}{3}, \ion{O}{4}, and \ion{N}{3}. In addition to these,the ground-state line
of [\ion{P}{2}] is also detected  for W28 and 3C391. 
Some of the other lines were detected in the younger and brighter 
supernova remnant RCW 103 (\cite{oliva}); upon reanalysis of these data
from the {\it ISO} archive, we find that the detected lines 
include the ground-state line of [\ion{P}{2}].
We can use these general principles to predict other bright lines, outside of
our observed spectral range. For example,
Ne is a very abundant element with lines in the 
mid-infrared; given that we detected ionized states of O, we expect
bright emission from [\ion{Ne}{2}] at 12.8 $\mu$m and perhaps
[\ion{Ne}{3}] at 15.5 $\mu$m.

\epsscale{0.8}
\plotone{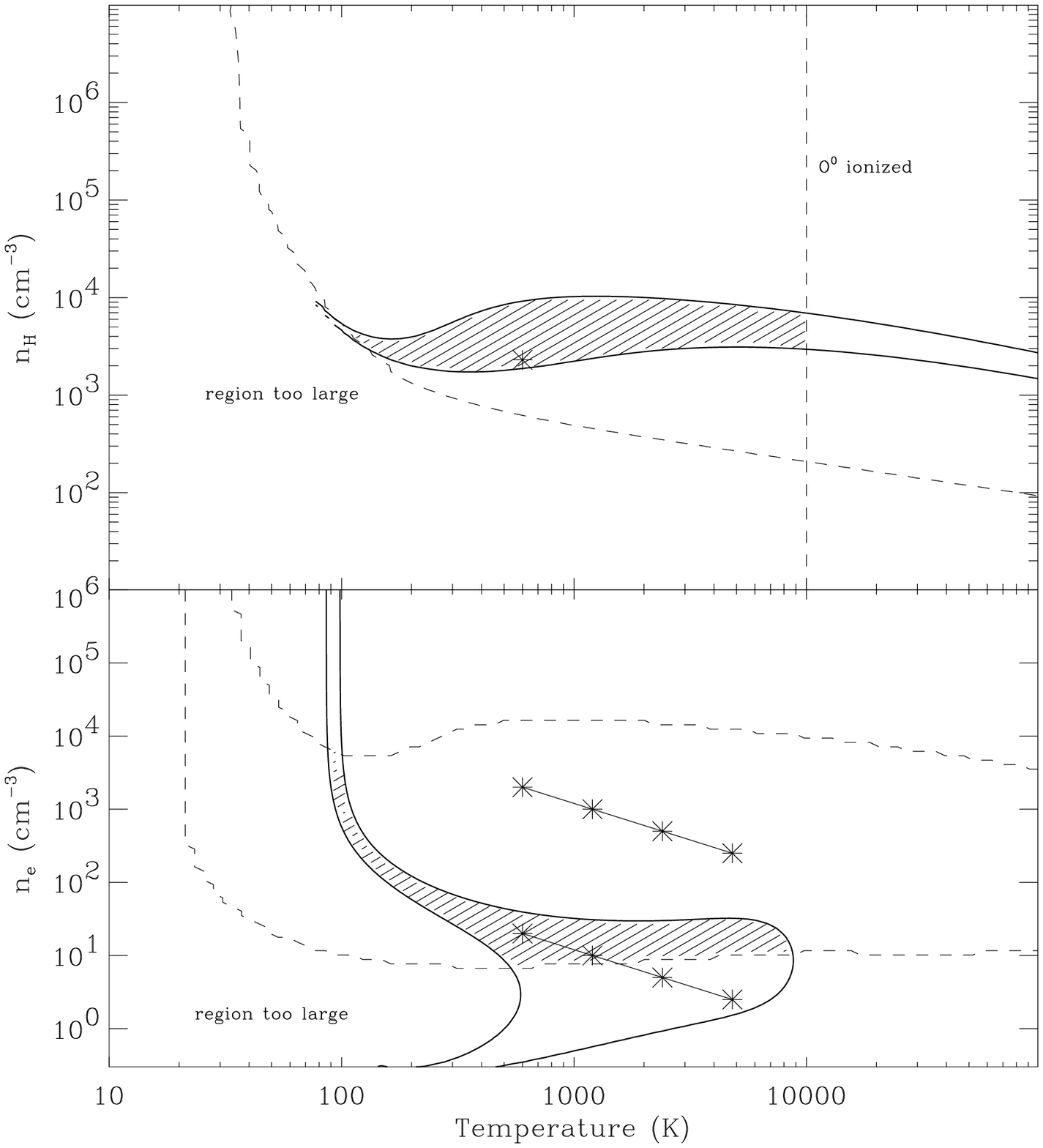}
\figcaption{\it Excitation diagrams for the [\ion{O}{1}] {\it (upper panel)} and
[\ion{O}{3}] {\it (lower panel)} lines.
The solid contours show the observed range of 145/63 $\mu$m {\it (upper panel)} 
and 52/88 $\mu$m {\it (lower panel)} line ratios from
Tabel~\ref{spectab}.\label{fig:excite}}
\begin{flushleft}{\it The dashed contour running
approximately from the upper left to lower right of each panel shows the
constraint placed on the density and temperature by requiring that the
emitting region is smaller than the entire supernova remnant, while still
producing lines as bright as observed. 
The vertical dashed line in the upper panel at $T=10^4$~K shows where O should
be collisionally ionized. The dashed contour 
near the top of the lower panel shows the constraint
that the \ion{O}{3} region be at least as hot and no more dense than
the \ion{O}{1} region. 
Hatched regions in both panels show the allowed solutions.
The asterisk in the upper panel shows the solution that we chose for 
detailed calculations. The upper-left asterisk in the lower
panel shows this same solution if the gas is fully ionized, and the asterisks
connect to it show the set of
solutions in thermal pressure equilibrium with it.
The lower set of asterisks shows the
same solutions assuming the ionization fraction in the \ion{O}{3}
region is 1\%.
}\end{flushleft}
\raggedright

\section{Abundances of the shocked gas\label{section:abundance}}

\subsection{Excitation of Oxygen}

Based on the arrangement of the energy levels of the ground-state atomic
fine structure, the ratios the upper and lower [\ion{O}{1}] and [\ion{O}{3}] lines
constrain the density and temperature of the emitting region, independent
of the geometry. 
The upper panel of Figure~\ref{fig:excite} shows the line ratio
predicted for [\ion{O}{1}] as a function of the total H density and
temperature. 
At each density and temperature, the 
level populations were calculated and the emergent intensity 
predicted using the escape probability (\cite{mh79}). 
The column density, $N_{\rm H}$, was adjusted iteratively for each model such that
the model predicts the correct brightness of the [\ion{O}{1}] 63 $\mu$m
line. Solid lines show the locus of models that predict the 
observed line ratios from Table~\ref{spectab}.
The locus of models for which the path length through the
emitting region, $z=N_{\rm H}/n_{\rm H}$, is equal to 3 pc (i.e. comparable to the
radius of the supernova remnant) is shown as a dashed contour, and
the region below this contour is labeled `region to large.'
Low-temperature and low-density models are not capable of producing the
observed line brightness within the confines of the observed region.
Another constraint is drawn as a vertical line at $T=10^4$ K, where
the neutral O atom would be collisionally ionized; higher temperature
models should therefore be excluded.
The lower panel of Figure~\ref{fig:excite} shows the line ratio
predicted for [\ion{O}{3}] as a function of density and temperature,
with each model normalized to predict the correct brightness of the [\ion{O}{3}] 
88 $\mu$m line.
The [\ion{O}{3}] and [\ion{O}{1}] emitting regions are not necessarily
coincident, so we consider the density, temperature, and ionization
in the two regions separately. One relative constraint between the two emitting
regions is practically guaranteed: the temperature will be at least as high,
the density will be at least as low (and the ionization will be 
at least as high) in the [\ion{O}{3}]
emitting region as compared to the [\ion{O}{1}] emitting region.
This rules out the very dense models for the [\ion{O}{3}] emitting region,
as indicated by the upper dashed curve in the lower panel of
Figure~\ref{fig:excite}.

To determine the physical conditions where the lines are produced, and to
determine the abundances of the elements in the shocked gas, we define
two emitting regions. Region M (molecular) is the [\ion{O}{1}] emitting region, with
density, temperature, and column density set to give the observed brightnesses
of the 63 and 145 $\mu$m lines. 
Region A (atomic) is the [\ion{O}{3}] emitting region, with
density, temperature, and column density set to give the observed brightnesses
of the 52 and 88 $\mu$m lines. Such a separation of emitting regions is
anticipated because we expect that
the neutral and doubly-ionized states of O would not
be co-spatial---although there could in principle be overlap,
if the gas is far from ionization equilibrium. 
Combining all of the constraints, the plausible range
of density and temperature for region M has
density $2000<n_{\rm H}^{(M)}$~(cm$^{-3}$)$<5000$ and
temperature $100<T^{(M)}$~(K)$<1000$ K.
For convenience, we have chosen a nominal solution of
$n_{\rm H}^{(M)}=3000$ cm$^{-3}$ and $T^{(M)}=600$ K, which is marked with
an asterisk ($\ast$) in Figure~\ref{fig:excite} and listed in Table~\ref{tab:cartoon},
for detailed calculations.

In region A, there is a wider range of solutions within
the constraints already mentioned. In particular, models with
constant $n_e=n_{\rm H}  x$, where $x$ is the fraction of H that is ionized,
produce the same excitation for the ions in region A.
If region A and region M were layers behind the same shock front, then 
we expect region A to have a comparable or somewhat higher pressure than
region M (lest the shock move backwards).
For the nominal solution for region M, the region A solutions
with the same pressure but density lower by factors of 1, 2, 4, and
8 are shown as two lines connecting asterisks in the lower panel of
Figure~\ref{fig:excite}. 
The upper line applies if region A is completely ionized, and the upper
left asterisk is just a direct copy of the region M solution.
It is evident that region A cannot be completely ionized {\it and}
be a shock layer upstream from region M: the region A pressure would be
so low that the shock would move backwards. 
The lower line applies if region A has an ionization fraction 
$x=10^{-2}$, and the upper-left asterisk of the lower line is
the region M solution with $n_e=n_H/100$.
This would be a very surprising result, because it means that O$^{++}$
coexists with a significant amount of H$^0$. Very low density
gas can be far from ionization equilibrium, with higher ionization states
coexisting with neutral states (\cite{nei}), but at the high densities required to
get the observed brightness of the 63 and 88 $\mu$m lines the ionization
should be reasonably close to equilibrium. For this reason, we consider
the [\ion{O}{3}] emitting region (region A) separate from the [\ion{O}{1}]
emitting region.
For detailed calculations, the region A model has 
density $n_e^{(A)}=8$ cm$^{-3}$ and temperature $T^{(A)}=3000$ K; the properties
are listed in Table~\ref{tab:cartoon}.
In region A, the abundance of O$^0$ is set to zero, 
and the abundance of O$^{++}$ and N$^{++}$
are set to the entire cosmic abundance of O and N, respectively.


\subsection{Predicted line brightnesses for different regions\label{sec:cartoon}}

The impact of a supernova blast wave with a molecular cloud leads to a 
complex interaction, with regions of different density reacting in 
very different ways. Based on the excitation of
[\ion{O}{1}] and [\ion{O}{3}] (discussed in the previous section),
and the excitation of H$_2$O and OH (\cite{rr98}), we identify
three types of post-shock gas and their infrared emitting lines. 
Figure~\ref{fig:cartoon} is a cartoon showing these shocks,
and Table~\ref{tab:cartoon} lists their properties.
The lowest-density pre-shock gas is region A
(for `atomic') with pre-shock density $n_0^{(A)}< 1$ cm$^{-3}$.
(If the pre-shock density of region A were higher than 1 cm$^{-3}$,
the post-shock density would be much higher than the 8 cm$^{-3}$ and
would be inconsistent with the 52/88 $\mu$m line ratio.)
The post-shock gas in region A gives rise to bright
\ion{H}{2} recombination lines and infrared and optical 
\ion{O}{3} and \ion{N}{3} (and other high ion) forbidden lines.
The moderate-density pre-shock gas is region M (for `molecular')
with pre-shock density $n_0^{(M)}\sim 10^2$ cm$^{-3}$. 
The post-shock gas in region M gives rise to bright \ion{O}{1} and
most of the other bright infrared forbidden lines we observed, as well
as some optical emission (which is, however, highly obscured).
The highest-density
pre-shock gas is in the dense clumps, with pre-shock density 
$n_0^{(C)}\sim 10^4$ cm$^{-3}$. The post-shock gas in region C gives rise
to bright H$_2$, OH, CO, CS, and H$_2$O emission, with a possible contribution
to the \ion{O}{1} emission. We do not mean to imply that there are three 
sharply-defined regions with fixed densities; we do suggest that the full range
of pre-shock densities is needed, such that no single region is consistent with
all of the observed spectral lines.

The shock velocities in the different regions are not well known, but reasonable
estimates can be made. If region A is the lowest-density pre-shock gas, then
it should represent the leading edge of the supernova remnant. 
The Sedov solutions for these remnants are consistent with the density of
region A for explosions with energy of order $10^{51}$ erg (\cite{rhothesis}).
Thus the shock velocity should be of order $2R/5t$, 
where $R$ is the radius  and $t$ is the age of the remnant. This works
out to $V_s^{(A)}\simeq 500$--600 km~s$^{-1}$ for the remnants considered in
this paper (\cite{rhothesis}). For region M, we have some direct evidence
of the shock velocity from the width of the \ion{O}{1} line (\S\ref{fabry})
that $V_s^{(M)}\simeq 100$ km~s$^{-1}$. This is also consistent with the
expanding H~I shell observed toward W~44 (\cite{kooheiles}). For region C, there
is some direct evidence of the shock velocity from the width of the 
millimeter-wave CO and CS lines (\cite{rr99}) that $V_s^{(C)}\simeq 30$ km~s$^{-1}$.
The shock ram pressure $p_{ram}\propto n_0 V_s^2$ should be comparable in 
these regions as they are all driven by the same blast wave.
Indeed we find $p_{ram}$ is progressively higher
for the denser regions, but the range of $p_{ram}$ is only a factor of 36.
A pressure enhancement in the denser regions has been theoretically 
justified by Chevalier (1999) as due to the shock into the denser
gas (regions M and C) being driven from the radiative shell (region A)
as opposed to being isolated shocks driven by the blast wave.

\plotone{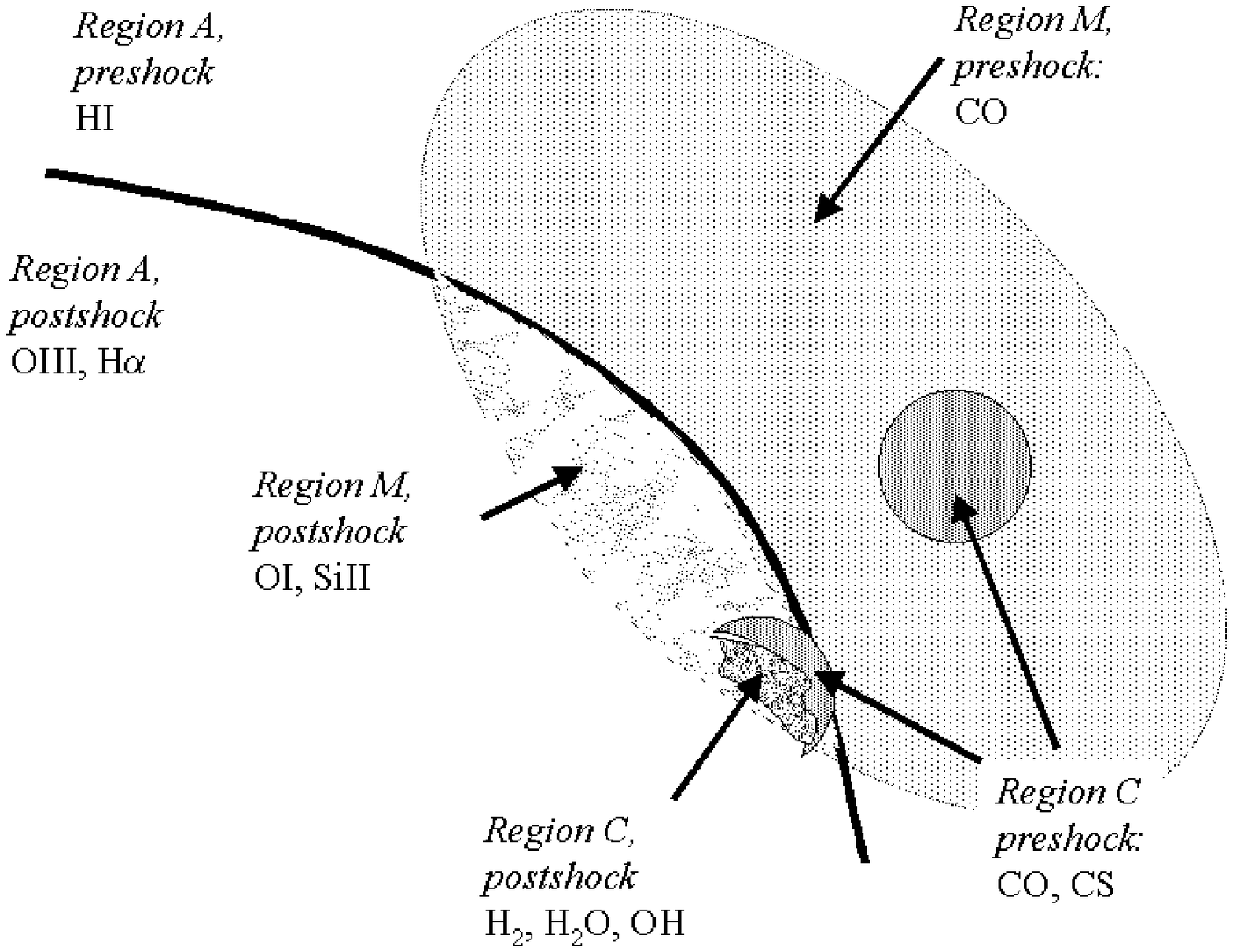}
\epsscale{1}
\figcaption{\it Cartoon illustrating the different pre- and post-shock regions
discussed in the paper. The physical properties of the regions are listed
in Table~\ref{tab:cartoon}. For each region, we also list an ion or 
molecule whose emission can be used to image the region.
\label{fig:cartoon}}

\begin{table}
\caption[]{Physical properties of the three post-shock emitting regions}\label{tab:cartoon} 
\begin{flushleft} 
\begin{tabular}{cccccrcl}
\hline
& \multicolumn{2}{c}{pre-shock} & ~ & \multicolumn{3}{c}{post-shock} & \\ \cline{2-3} \cline{5-7}
\multicolumn{1}{c}{Region} & \multicolumn{1}{c}{$n_0$} & \multicolumn{1}{c}{$V_s$} & &
\multicolumn{1}{c}{$n$} & \multicolumn{1}{c}{$T$} & 
\multicolumn{1}{c}{$N$} & \multicolumn{1}{l}{Infrared Coolants} \\
\multicolumn{1}{c}{} & \multicolumn{1}{c}{(cm$^{-3}$)} & \multicolumn{1}{c}{(km~s$^{-1}$)} & &
\multicolumn{1}{c}{(cm$^{-3}$)} & \multicolumn{1}{c}{(K)} & 
\multicolumn{1}{c}{(cm$^{-2}$)} \\
\hline \hline
A (atomic) &  $<1$  & 500 & & 8              & 3000 & $10^{20}$ & 
	\ion{O}{3}, \ion{N}{3} \\
M (molecular) & $10^2$ & 100 & & $3\times 10^3$ &  600 & $10^{21}$ & 
	\ion{O}{1}, \ion{Si}{2},\ion{Fe}{2}, \ion{N}{2}, H$_2$S(9) \\
C (clump) & $10^4$ & 20 & & $4\times 10^5$ &  200 & $10^{22}$ & 
	CO, H$_2$O, OH, H$_2$S(3) \\
\hline
\end{tabular} 
\end{flushleft} 
\end{table}  

\raggedright

In Chevalier's (1999) paper, two regimes of density were considered
for the pre-shock gas; these densities correspond roughly to our regions A and M
(cf. Table~\ref{tab:cartoon}).
Chevalier explained the [\ion{O}{1}] 63 $\mu$m luminosity of the remnant using
only relatively low-density shocks into region A. Based on our new results for
the line ratio and excitation of \ion{O}{1} and the high-resolution spectrum
of the 63 $\mu$m line showing a width of only 100 km~s$^{-1}$, 
we suspect that the 63 $\mu$m arises in relatively denser regions, 
like region M.
Indeed, it is impossible for a low-density emitting region to produce 
63 $\mu$m lines as bright as we observed, within the confines of the supernova
remnant. In the model of Hollenbach \& McKee (1989), the 63 $\mu$m line
brightness is proportional to $n_0 V_s$. For the cartoon model parameters
in Table~\ref{tab:cartoon}, it is clear that the denser regions will dominate
the line brightness if they fill an appreciable portion of the beam. In
region C, however, the O is probably largely tied up in molecular form;
furthermore, region C is somewhat beam diluted in the {\it ISO} LWS beam.
Therefore, we believe that the 63 $\mu$m lines presented here arise from shocks into
the moderate-density molecular gas of region M.

\begin{table}
\caption[]{Model predictions for fine-structure lines}\label{tab:abundance} 
\begin{flushleft} 
\begin{tabular}{lrrrrcc}
\hline
\multicolumn{1}{c}{Ion} & 
\multicolumn{1}{c}{Abundance} & 
\multicolumn{1}{c}{line wavelength} & 
\multicolumn{1}{c}{Atomic} & 
\multicolumn{1}{c}{Observed} & 
\multicolumn{2}{c}{Observed/Predicted} \\ \cline{6-7}
\multicolumn{1}{c}{} &  
\multicolumn{1}{c}{} &  
\multicolumn{1}{c}{($\mu$m)} & 
\multicolumn{1}{c}{Data$^{a}$} & 
\multicolumn{1}{c}{(erg~cm$^{-2}$~s$^{-1}$~sr$^{-1}$)} & 
\multicolumn{1}{c}{layer M$^{b}$} &
\multicolumn{1}{c}{layer A$^{c}$} 
\\ \hline \hline
\ion{C}{2}  &  -3.44 & 157.7 &  1   & $5.8\times 10^{-4}$ &   1.2 &   ... \\
\ion{N}{2}  &  -3.95 & 121.9 &  1   & $2.7\times 10^{-4}$ &   0.3 &   10  \\
\ion{N}{3}  &  -3.95 &  57.3 &  2,3 & $9.7\times 10^{-3}$ &   ... &   0.3 \\
\ion{O}{1}  &  -3.07 &  63.2 &  4,5 & $1.3\times 10^{-3}$ &   1.0 &   ... \\
\ion{O}{1}  &  -3.07 & 145.5 &  4,5 & $9.1\times 10^{-5}$ &   1.0 &   ... \\
\ion{O}{3}  &  -3.07 &  88.4 &  1   & $5.2\times 10^{-3}$ &   ... &   1.0 \\
\ion{O}{3}  &  -3.07 &  51.8 &  1   & $2.8\times 10^{-3}$ &   ... &   0.9 \\
\ion{Si}{2} &  -4.45 &  34.8 &  5   & $3.0\times 10^{-3}$ &   0.5 &   10  \\
\ion{P}{2}  &  -6.55 &  60.6 &  6,7 & $8.6\times 10^{-6}$ &   1.1 &   ... \\
\ion{Fe}{2} &  -4.33 &  26.0 &  8   & $2.7\times 10^{-4}$ &   0.3 &   10 \\ \hline
\end{tabular} 
\end{flushleft} 
\noindent { $^{a}$References for atomic data:
1 (\cite{spitzer}), 2 (\cite{froese}), 3 (\cite{blum}),
4 (\cite{launay}), 5 (\cite{HM89}), 6 (\cite{mendoza}), 
7 (\cite{piicoll}), 8 (\cite{nussbaumer})}

\noindent { $^{b}$layer M model parameters: $n=3000$ cm$^{-3}$, $T=600$ K,
	\ion{O}{3} and \ion{N}{3} abundances set to zero, column density
	normalized to produce observed [\ion{O}{1}] lines}

\noindent { $^{c}$layer A model parameters: $n_e=8$ cm$^{-3}$, $T=3000$ K, \ion{O}{1}
		abundance set to zero, column density normalized to produce
		observed [\ion{O}{3}] lines}
\end{table}  

\raggedright

For regions A and M, we calculated the model brightness of all of the 
lines that we observed with {\it ISO}. 
Each element was assumed to have its 
entire solar abundance in the gas phase and in the ionization state that
gives rise to the line, with the following exceptions. In region M,
oxygen was assumed to be neutral (as we observed), 
and nitrogen was assumed to be singly ionized (since the only
detected line was from \ion{N}{2}, not \ion{N}{1}) in order to obtain an upper
limit to the predicted \ion{N}{2} brightness..
In region A, oxygen was assumed to be ionized. 
Our purpose in these calculations is to measure the richness or depletion relative to
solar abundances. Our calculations do not yield absolute abundances, because in reality
the total abundance of each element is actually divided among the ionization states 
that are present. In the next section, we go through the list of elements and discuss their
expected ionization state and the inferred total abundances in the post-shock gas.
Table~\ref{tab:abundance} shows the brightness predictions for regions A and M.
We can conclude the following, without considering the ionization balance.
First, it is possible to reproduce all of the observed fine-structure lines, except
\ion{N}{3} and \ion{O}{3} using region M alone. (In fact, it is 
possible to get the \ion{N}{3} and \ion{O}{3} lines, if we could accept the
multiply ionized species existing in region M, where H is mostly neutral.)
Therefore, we expect to be able to derive abundances in the `region M'
post-shock gas. For region C, we discuss details in \S\ref{sec:h2} below.

\subsection{Post-shock gas abundances}

We now step through each element and derive the total
abundance of that element in the post-shock gas.
Table~\ref{tab:abundancetwo} summarizes the abundances inferred for the
post-shock gas. For each element $X$, $[X/H]_\odot$ is the solar abundance
(\cite{anders}) and $[X/H]_{obs}$ is the total abundance of that element
from our observations and model.

\begin{table}
\caption[]{Abundances}\label{tab:abundancetwo} 
\begin{flushleft} 
\begin{tabular}{lrr}
\hline
\multicolumn{1}{c}{Element} & \multicolumn{1}{c}{$\log{[X/H]_\odot}$} & 
\multicolumn{1}{c}{$[X/H]_{obs}/[X/H]_\odot$}
\\ \hline \hline
C  & -3.44 & $< 1.1$ \\
N  & -3.95 & $>0.3$$^a$ \\
O  & -3.07 & 1 \\
Si & -4.45 & 0.5 \\
P & -6.55 & 1.1 \\
Fe & -4.33 & 0.3 \\
\hline
\end{tabular} 
\end{flushleft} 
\noindent $^a$ abundance of N is a lower limit because the dominant ionization state
(\ion{N}{1}) was not observed\par
\end{table}  

\raggedright

\noindent{\it Carbon---}The only carbon line we observed was from \ion{C}{2},
which is likely to be the dominant ionization state in region M. The observed
line brightness compared to the region M prediction (Table~\ref{tab:abundance})
requires essentially all of the carbon to be in the gas phase. However, a
significant fraction of the 157 $\mu$m line could come from line-of-sight gas,
so this result is inconclusive. The carbon abundance in the post-shock gas could
be better addressed if we knew the brightness of the line just outside
of the remnant. Using our current knowledge, we can only say that the 
{\it ISO} observations are consistent with complete destruction of the
C-bearing dust grains if all of the emission comes from the remnant, 
or partial or no destruction if less than half of the emission comes 
from the remnant. In Table~\ref{tab:abundancetwo} we list the inferred
depletion of carbon in the post-shock gas assuming {\it all} of the 157 $\mu$m
line comes from the remnant; this yields an upper limit to the 
gas-phase abundance, and therefore lower limit to the depletion.


\noindent{\it Nitrogen---}We observed nitrogen in its singly- and doubly-ionized
states. The dominant state in region M should be neutral, because the ionization
potential of \ion{N}{1} is greater than 13.6 eV. Comparing the observed
to predicted brightness (Table~\ref{tab:abundance}),
we require only about 1/3 of the nitrogen to be \ion{N}{2},
which is reasonable. 
When all nitrogen is assumed to be singly ionized, the
observed to predicted brightness ratio of [\ion{N}{2}] is only 0.3
(Table~\ref{tab:abundance}),  which implies either nitrogen has lower 
abundance or the predicted \ion{N}{2} is too high. Since the former is
unlikely, the ratio implies that only about 1/3 of the nitrogen is
\ion{N}{2} and the rest is neutral.
In region A, which was normalized to give the
correct 88 $\mu$m line brightness, we require about 1/3 of the \ion{N}{3}, which
is also plausible. In Table~\ref{tab:abundancetwo} we list the depletion of
nitrogen assuming all of the nitrogen is ionized; this is a lower limit
to the gas-phase abundance, and an upper limit
to the depletion, because it neglects the neutral state.

\noindent{\it Oxygen---}The column density of our model was adjusted to match 
the observed brightness of the oxygen lines, so most of our measurements are,
effectively, relative to the O abundance. We cannot easily determine
the total abundance of O because we lack an accurate measurement of the 
total column density of H.

\noindent{\it Silicon and Iron---}We find that a significant fraction of 
the total abundance of Si and Fe, which 
are normally highly depleted  into dust grains in the interstellar medium, 
is required to produce the observed line brightness. 
Both Si and Fe were observed in their dominant ionization state in region M.
Region A does not have enough column density to contribute significantly
to the observed line brightness. (Recall that the region A column density
is set by the [\ion{O}{3}] line brightness, so increasing its column
density would make too much [\ion{O}{3}] emission.)
For region M, we find that about 1/2 of the Si and 1/4 of the Fe are required to 
be in the gas phase.  These abundances of Fe and Si are
interestingly comparable to those of gas-phase abundances in
intermediate-velocity interstellar clouds which are recently measured 
using Goddard High Resolution Spectrograph, suggesting that resilient
cores of grains are not easily destroyed in shocks (\cite{fitzpatrick}).
We will discuss the implications for grain destruction
more in \S\ref{graindest} below.

\epsscale{0.6}
\plotone{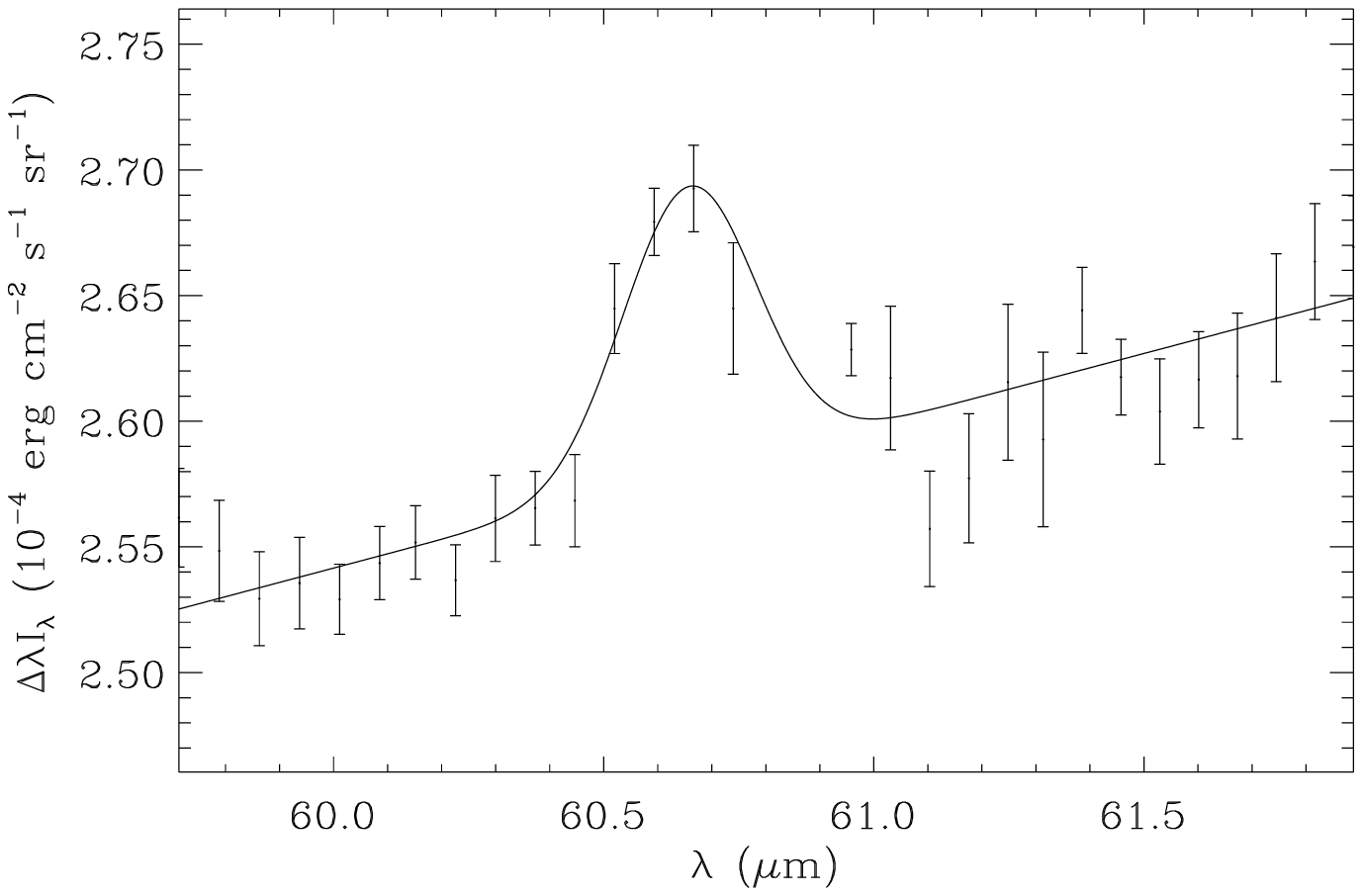}
\epsscale{1}
\figcaption{\it Spectrum of the \ion{P}{2} 60.6 $\mu$m line toward 3c~391:BML. The
solid curve is a Gaussian fit to the line with width fixed to the LWS grating
spectral resolution of $\Delta\lambda=0.29$ $\mu$m.
\label{piispec}}
\raggedright

\noindent{\it Phosphorus---}The abundance of P is known for only a few lines of sight
in the interstellar medium. Our abundance estimate in the post-shock gas
uses the observed brightness of the 60.6 $\mu$m line, for which this
is the first reported detection (at the time of writing). 
Figure~\ref{piispec} shows the spectrum of  this newly-detected line.
We used the radiative decay rates are as given by Mendoza \& Zeippen (1982),
and we use the collision strength $\Omega(^3{\rm P}_1\rightarrow 0)=1.46$
(\cite{piicoll}).
The P abundance relative to O is very well constrained because the energy
levels of the two observed ions are similar, and both elements were
observed in their dominant ionization state. We find that the observed line
brightness is consistent with solar abundance of P.
Along the line of sight toward $\zeta$ Oph, it appears that about half of
interstellar P is normally locked in grains (\cite{sembachsavage}).
Our observation suggests that some or all of the P-bearing solid material
was vaporized by the strong shock wave, 
as we find for Fe and Si (see \S\ref{graindest}).

\section{Shock-excited molecules\label{sec:h2}}

\subsection{H$_2$}

The S(3) and S(9) lines of H$_2$ were both detected from each of the 
supernova remnants we observed, and the line brightnesses are listed
in Table~\ref{swstab}.
These H$_2$ lines arise from shocks in the molecular gas,
like the CO, OH, and H$_2$O lines (\cite{rr98}). Using our cartoon
model, we can now localize which region(s) give rise to which lines and
further constrain the properties of the dense clumps (region C).
First, for region M, if we assume that the H$_2$ rotational lines are
collisionally excited with rotational excitation temperature equal to the
kinetic temperature, we predict that the brightest H$_2$ line is the S(3) line.
The brightness of the S(3) line from region M, assuming all of the H is
in H$_2$, is $2\times 10^{-4}$ erg~cm$^{-2}$~s$^{-1}$~sr$^{-1}$. This is
in very good agreement with the observed line brightness, suggesting that
the H is largely molecular behind the region M shocks. The column
density of H$_2$ is just enough that H$_2$ molecules are self-shielded
from dissociating photons in the interstellar radiation field (\cite{h2form}).

We cannot explain the brightness of the higher H$_2$ S(9) rotational
line, or the higher-lying ro-vibrational lines in the near-infrared, 
using the region M shocks. These lines most likely arise in 
molecular regions that are much warmer than the recombined molecules
behind a fast, dissociative shock. Instead, the S(3) lines probably
arise from non-dissociative shocks into denser gas (discussed below).

In a previous paper (\cite{rr98}), we derived the physical properties
of region C using the excitation of CO, OH, and H$_2$O. In that paper,
we found the puzzling result that abundance of H$_2$O was
lower than expected in a region with `warm' chemistry, where O is
efficiently converted into molecules and OH is converted into H$_2$O.
Two papers have shown that OH can be formed from dissociation of H$_2$O
(\cite{lockett}, \cite{wardle}), so the abundance OH is no longer 
such a puzzle.
We now reexamine the H$_2$O abundance relative to H$_2$. 
The column density of H$_2$O was
found to be $\sim 3\times 10^{17}$ cm$^{-2}$ using the far-infrared line
brightnesses. The column density of H$_2$ was taken from the volume density
times path length, yielding $\sim 8\times 10^{23}$ cm$^{-2}$. 
We suggested this column density was consistent with the brightness of the 
H$_2$ S(3) line, but we now suspect that much of the S(3) line
arises from region M and not the dense shock of region C.
If this is true, then the H$_2$ column density of region C could be
much lower than we estimated before. 
Let us start with some assumptions about the molecular abundances
and infer the H$_2$ abundance. First, CO:OH:H$_2$O is 100:1:15 from our 
observations (\cite{rr98}).
Assume that the C and O are largely in molecules. If a fraction
$f_{CO}$ of the C is locked in CO, then the abundance ratio of H$_2$O/H$_2$
is $X_{\rm O} f_{CO}$, where $X_{\rm O}$ is the abundance of O. 
The column density of H$_2$ is then
$N({\rm H}_2)^{(C)}=3\times 10^{21} f_{CO}^{-1}$. Such a column density
for region C seems reasonable, but it is lower than our previous estimate
based on the observed size of the CO emitting region in the millimeter
observations (\cite{rr99}). This could be explained by a low filling factor
for the dense gas, of order $0.007 f_{CO}^{-1}$ in the {\it ISO} SWS beam.
Taking $f_{CO}=0.5$ as an example, the size of the emitting region would be 
of order $1^{\prime\prime}$, and it would be highly structured in future, 
higher-angular-resolution observations. 
If true, this new interpretation would change our estimate of the mass of 
the 3C~391:BML
post-shock gas, making the self-gravity of the shocked gas negligible,
and weakening our argument that the dense clump might imminently 
collapse to form stars (\cite{rr99}).
Table~\ref{tab:cartoon} shows
the cartoon model for region C assuming $f_{CO}=0.5$.

The H$_2$ 0-0 S(9) line cannot be assigned to any of the emitting regions
that we have defined so far. It's energy level is too high above the ground
state to be excited in the cooling region where the far-infrared lines
are produced. It is unlikely that the S(9) line can be produced in the
region M shock, even upstream from the cooling region, because the H$_2$
is probably still dissociated. Thus we suspect that the S(9) line is
produced upstream from the region C shock. To get the observed line S(9)
brightness, in an emitting region with rotational excitation temperature
of order 1500 K, a column density of order $10^{19}$ cm$^{-2}$ is needed,
which could easily fit in the upstream region considering the cooling
region has a column density 600 times larger. The 1500 K excitation 
temperature seems high, but high excitation is predicted by
shock models (\cite{DRD}) and is required to explain observed
line ratios in IC~443 (\cite{rho00}, \cite{richter}).
Future observations should be able to distinguish which shocks produce
the S(3) and S(9) lines, by measuring the {\it width} of the lines.
According to our cartoon model, much of the S(3) line comes from region M and
would have a width of 100 km~s$^{-1}$. The S(9) line comes upstream from
region C and would have a width of 30 km~s$^{-1}$. The line profiles of
progressively higher excitation lines would evolve from wide to narrow
lines.

\epsscale{0.7}
\plotone{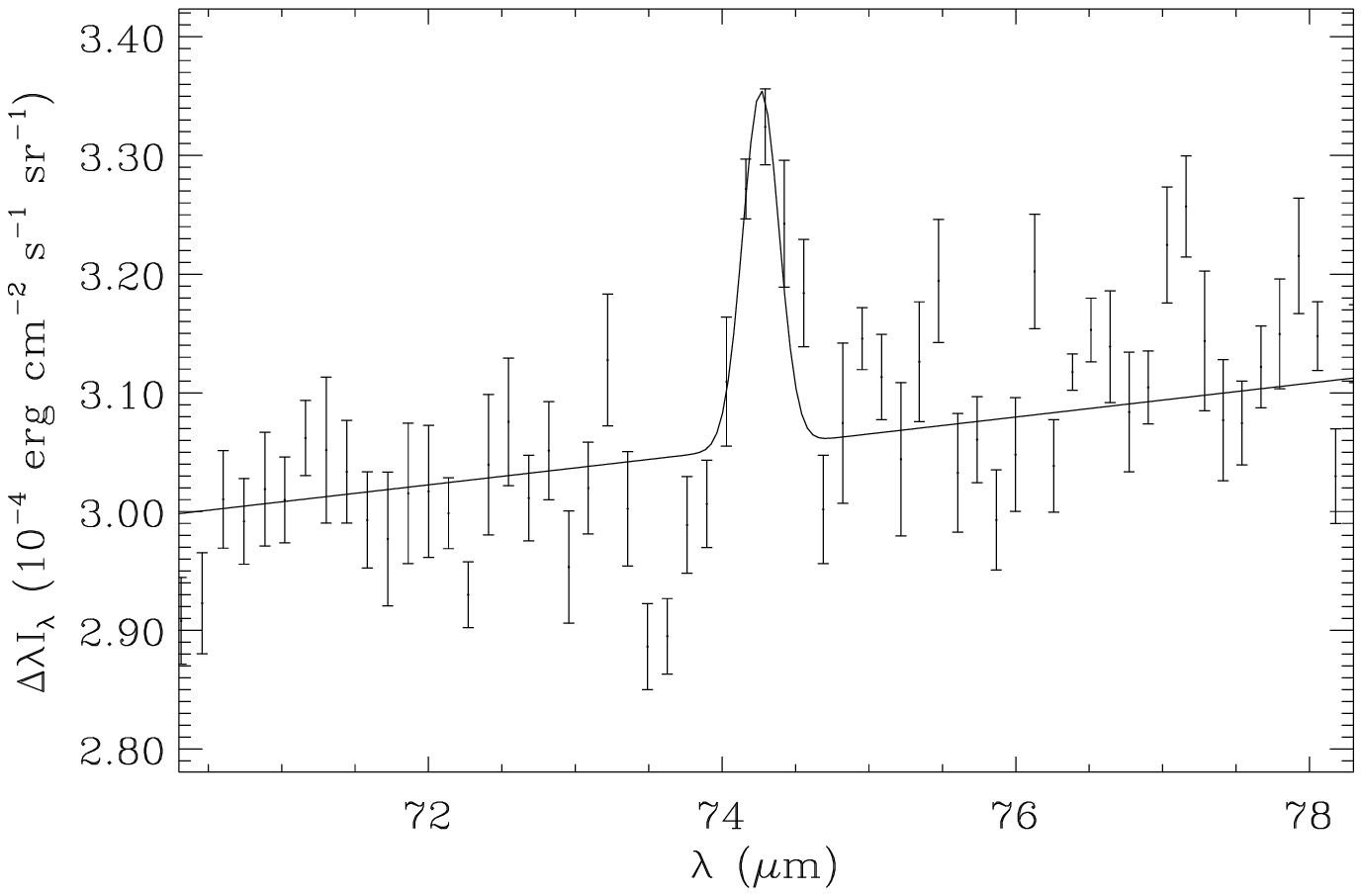}
\epsscale{1}
\figcaption{\it Spectrum of an unidentified line toward 3C~391:BML. The
solid curve is a Gaussian fit to the line with width fixed to the LWS grating
spectral resolution of $\Delta\lambda=0.29$ $\mu$m.
\label{seventyfour}}
\raggedright

\subsection{Unidentified line at 74.3 $\mu$m}

One relatively bright line remains unidentified. 
Figure~\ref{seventyfour} shows the spectrum of the 74.3 $\mu$m line
for 3C~391:BML. The line was present
in our complete LWS spectra of W~28, W~44, and 3C~391, and it
has also been seen in the remnant RCW~103 (\cite{oliva}) and the planetary
nebula NGC 7027 (\cite{liu}). Table~\ref{unident} lists
the brightnesses and central line wavelengths for our remnants.
The wavelengths are all consistent with 74.26 $\mu$m. 
All of the spectra that contain this line contain also contain
both atomic fine structure and
molecular lines. Based on our list of plausible lines
compiled above in Tables~\ref{tab:periodic} and~\ref{tab:periodictwo}, 
and a careful inspection of all atomic fine structure
lines, we can find no candidate atomic or ionic source for the line.
(The only fine structure line that matches the observed wavelength
is a transition among high energy levels of [\ion{Ti}{3}], which
is extremely unlikely based both on abundance and energetics.)
Therefore, we suspect this line is molecular in origin. Despite
searching the JPL spectral line database (\cite{pickett}), we cannot
identify the origin of this line, which is apparently caused by
an unusual molecule or meta-stable state not yet found in laboratories
or planetary atmospheres. 

\begin{table}
\caption[]{Unidentified line}\label{unident} 
\begin{flushleft} 
\begin{tabular}{lcc} 
\hline
\multicolumn{1}{c}{source} & \multicolumn{1}{c}{wavelength} & 
\multicolumn{1}{c}{ Line Brightness}\\
& \multicolumn{1}{c}{($\mu$m)}& \multicolumn{1}{c}{($10^{-4}$ erg~s$^{-1}$cm$^{-2}$sr$^{-1}$)}\\
\hline \hline
3C~391:BML & $74.26\pm 0.03$ & $0.31\pm 0.05$ \\
W~44       & $74.32\pm 0.06$ & $0.11\pm 0.03$ \\
W~28       & $74.25\pm 0.03$ & $0.26\pm 0.04$ \\
\hline
\end{tabular} 
\end{flushleft} 
\end{table}  

\raggedright

\section{Continuum and Grain Destruction\label{section:continuum}}

\subsection{Continuum spectrum}

The moderate-resolution far-infrared spectra are dominated by a bright continuum
underlying the spectral lines. Because the remnants we discuss here are in the galactic
plane, much of this continuum emission is likely to be unrelated to the remnant,
produced instead by clouds along the line of sight
both in front of and behind the remnant, as well as the molecular cloud close to
the remnant. We have attempted to extract the remnant contribution to the continuum
emission by searching for distinct spatial and spectral variations. In this
paper we present a spectral separation of the remnant emission from the unrelated
emission; in paper 1 and in future work, we use the reference positions and
spatial correlations with the \ion{O}{1} line 
brightness\footnote{
We would like to clarify an issue from paper 1.
Despite a statement by Oliva et al. (1999a) that our continuum
for W~44 is 10 times too bright compared to {\it IRAS} data,
the continuum brightness of the remnant as observed with {\it ISO}
appears real. The {\it IRAS} data suffer from lower angular resolution,
which makes it extremely difficult to separate the remnant from
unrelated galactic emission.
The continuum emission will be discussed in more
detail in a forthcoming paper.}.

\epsscale{0.69}
\plotone{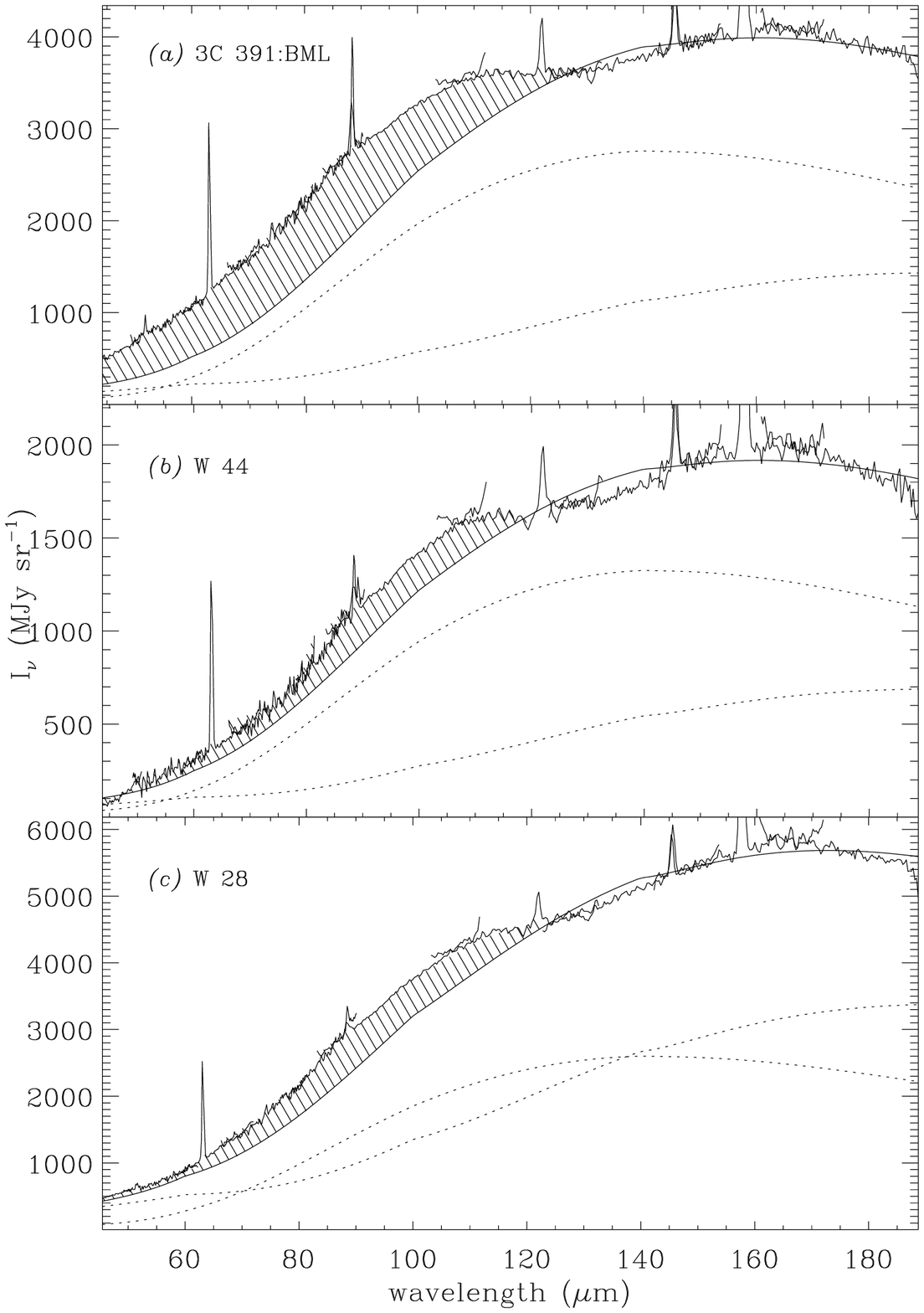}
\epsscale{1}
\figcaption{\it Far-infrared continuum spectra toward molecular shocks in
{\it (a)} W~28,
{\it (b)} W~44, and
{\it (c)} 3C~391.
The jagged curve shows the observed brightness.
The solid curve shows the spectral model for unrelated dust along the line
of sight. This model consists of two components of dust associated with
atomic gas ($T_d\sim 20$ K) and molecular gas $T_d\sim 13$ K); each component
is shown as a dotted curve.
The
hatched region shows the excess emission that could be the continuum emission from
the remnants. The two dotted curves show the individual contributions of unrelated
atomic and molecular gas.
\label{contspec}}

\raggedright

The continuum spectrum of 3C~391:BML is shown in Figure~\ref{contspec}. Even by
inspection, it is evident that the spectrum cannot be fitted with a modified
blackbody: there is a significant positive excess at wavelengths shorter than
120 $\mu$m. It is not surprising for the continuum spectrum to require 
multiple components. In the {\it COBE} FIRAS
observations of the galactic plane in a $7^\circ$ beam,
the spectrum of the inner galaxy at wavelengths of 100--200 $\mu$m has
contributions from two components: 20 K dust associated with atomic gas,
and 13 K dust associated with molecular gas (\cite{reachfiras}, \cite{lagachecold}). 
At wavelengths shorter than 100 $\mu$m, continuing all the way to the PAH
features in the mid-infrared, emission from small,
transiently-heated dust grains becomes important (\cite{dbp}).
For our supernova remnant observations in an $80^{\prime\prime}$ beam, we
expect all components to be brighter than in the {\it COBE}/FIRAS data, because 
(1) the atomic gas column density peaks sharply in the galactic plane, 
(2) the lines of sight pass directly through giant molecular clouds, 
and (3) the emission from the remnants themselves are completely beam-diluted
in the {\it COBE} FIRAS beam. 

To estimate the remnant contribution to the spectrum, we used a nominal
spectrum for dust in the local interstellar medium, which big grains,
at a temperature of 17.5 K, and small grains with a brightness
that matches the {\it COBE} Diffuse Infrared Background Experiment
(\cite{reachboul} and references therein). 
We use two components, for the atomic and molecular gas,
with radiation fields equal to $\chi$ times the local interstellar radiation field.
For each component, the temperature of big grains scales as $\chi^\frac{1}{6}$, and the
brightness of the small grain emission scales as $\chi$.
The atomic regions were assumed to have $\chi=2$ (to account for the enhanced
radiation field in the inner Galaxy), and the molecular regions were assumed to
have $\chi=0.5$ (to account for extinction of the ultraviolet radiation field
inside molecular clouds). The scale of the model and the
ratio of the two components was set to match the far-infrared spectrum
at wavelengths longward of 120 $\mu$m.

The continuum spectral
models are shown for each remnant in Fig.~\ref{contspec}.
The models with a single diffuse ISM component
cannot explain the brightness of the continuum at wavelengths shortward
of 110 $\mu$m; there is a clear excess at 40--110 $\mu$m.
The amount of excess emission depends rather sensitively on the assumed
spectrum of the unrelated interstellar material, so the present results 
must be used with caution. There is a good correlation of the 74 $\mu$m 
continuum with the \ion{O}{1} line brightness (paper 1), supporting our claim that
a significant portion of the 40--110 $\mu$m emission is from the remnants.
There is also a tentative detection of 3C~391 at 60 $\mu$m using {\it IRAS} data
(\cite{arendtcatalog}).
This excess emission has a color temperature of $\sim 40$ K, corresponding to
an enhancement of the radiation field by more than 2 orders of magnitude.
Such a high radiation does not occur in the diffuse interstellar medium
over many-pc scales such as observed here. Instead, we must be seeing dust
that is either physically different from that in the diffuse interstellar
medium, or dust that is excited locally, for example behind the shock fronts.
The optical depth of the remnant emission at 100 $\mu$m is 
$\tau\sim 2\times 10^{-4}$, and the total surface brightness of
the far-infrared continuum, assuming a $\nu^2 B_\nu(42\,{\rm K})$ spectrum,
is $2\times 10^{-2}$ erg~s$^{-1}$~cm$^{-2}$~sr$^{-1}$. Thus the infrared
continuum is significantly brighter than any of the individual spectral
lines.
The luminosity of the excess emission, within the single LWS beam that
we observed for each remnant, ranges from
$\sim 600 L_\odot$ for 3C~391:BML to $\sim 10 L_\odot$ for W~44 and W~28; 
however, these numbers are rather uncertain. 
Using the optical depth required to produce the excess emission,
the opacity per unit mass for `normal' interstellar dust (\cite{drainelee}),
the mass of shock-excited dust within the single LWS beam that we
observed for 3C~391:BML, is $\sim 1 M_\odot$.

\subsection{Grain destruction\label{graindest}}

In order for the Fe and Si emission lines to be as bright as observed,
at least part of the refractory dust mass must have been vaporized. 
In the cool diffuse cloud toward
$\zeta$ Oph, 95\% of Si and $>99$\% of Fe are locked in dust grains
(\cite{sembachsavage}). 
Using our single-slab model, we found that 
50\% of the Si and 30\% of the Fe are required to be in the post-shock {\it gas},
which would mean that about half of the dust mass is destroyed by the shock.
The single-slab model will, however, give less accurate measures of the 
abundances for Fe and Si relative to O (for which the model parameters
were determined), because the Fe and Si lines we observed are from higher
energy levels than those of O. There is likely to be a stratified post-shock
emitting region, with somewhat warmer regions producing some of the Si and Fe
lines. Thus, the exact fraction of refractory dust mass that was vaporized
is highly uncertain. We have experimented with a multiple-slab model, using
the approximate density and temperature profiles behind a 100 km~s$^{-1}$ J-shock
into $10^{3}$ cm$^{-3}$ gas (\cite{HM89}), to see how much of the observed 
35.2 and 26 $\mu$m line 
brightnesses could be produced by warmer layers closer to the shock. 
The multiple slab model predicts relatively brighter 26 $\mu$m
lines for a given column density, by up to a factor of two in extreme cases.
Thus it is possible that
the gas-phase abundance of Fe was overestimated by up to a factor of 2
using the single-slab model, meaning that as little as $15$\% of the Fe is
required to produce the observed lines. An identical result is obtained for Si.
Even with the lower gas-phase abundance of Si and Fe,
we can conclude that a non-trivial fraction
of the Fe-bearing dust mass was vaporized. 
It is essentially impossible to produce
all of the observed brightness of the 26 $\mu$m line from
normally-depleted pre-shock interstellar matter. 

Using the observed traces of the vaporized and surviving dust grains, 
we can work backwards to estimate the fraction of dust that was destroyed. 
The total mass of Si and Fe vapor
in 3C~391:BML is about 0.11 $M_\odot$, within the LWS beam. Based on the dust-phase
elemental abundances in a cold interstellar cloud,
the total mass of elements (mainly O, Fe, C, Si, and Mg) in dust is 3.6--5.0
times the mass of Si and Fe in dust, with the lower number coming from B-star
reference abundances and the higher number coming from Solar reference 
abundances (\cite{sembachsavage}).
Therefore, we infer that about 0.5 $M_\odot$ of dust was vaporized by shocks
into 3C~391:BML. For comparison, our far-infrared conntinuum observations
indicated at least 1 $M_\odot$ of dust remaining in solid form after the
shock. Therefore, we estimate that approximately 1/3 of the dust mass was
vaporized by the shocks.

How does our inferred fraction of dust vaporization compare to theoretical
predictions? Jones, Tielens, \& Hollenbach (1996) have calculated the
destruction probabilities for dust grains in shock fronts with a range of
shock velocities and pre-shock densities. Their models included 
sputtering and shattering: sputtering removes material from the grain via
collisions with the hot gas, while shattering destroys large grains and
creates small grains from them. The models for $v_s=100$ km~s$^{-1}$ are
most relevant for the shocks we discuss in this paper. Clearly there are
faster shocks that create the X-ray-emitting plasma, and slower shocks that
create the molecular line emission. But the ionic lines from which we
determined the abundances arise from a layer with a post-shock
density of order $10^3$ cm$^{-3}$ (\S\ref{section:abundance}) into which shocks 
with a velocity of about 100 km~s$^{-1}$ have been driven (\S\ref{fabry}).
The model of Jones et al. (1996) with parameters most similar to our
present case has pre-shock density $n_0=25$ cm$^{-3}$. While their models
do not extend to higher $n_0$, the dependence on $n_0$ is surprisingly mild,
with a slight trend toward {\it more} dust destruction for denser shocks.
The models of Jones et al. (1996) predict that about 37\% of silicate
dust and 12\% of graphite dust are destroyed, mostly via non-thermal sputtering.
These predictions are in excellent agreement with our results.

\section{Nature of the shock fronts powering the emission lines}

There are at least two significant theoretical models that predict
the structure (density and temperature) and emission behind
radiative shocks comparable to those that seem to be occurring 
in the molecular supernova remnants we observed. One model,
by Hollenbach \& McKee (1989; hereafter HM89), includes an
extensive network of chemical reactions and detailed
treatment of the dynamics of the post-shock gas (from \cite{mh79}).
The other model, by Hartigan, Raymond, \& Hartmann (1987; hereafter
HRH87), includes more ionization states and more ultraviolet and
optical lines. We will use the HM89 model as a benchmark, and
we will use the [\ion{O}{1}] 63 $\mu$m line as the normalizing factor.
In the HM89 model, the 63 $\mu$m line is a significant coolant and should
reliably indicate the total energy passing through the shock. 
Based on the observed width of the line (\S\ref{fabry}), we
consider the models with shock velocity $V_s=100$ km~s$^{-1}$.
The HM89 model at that velocity that matches the observed [\ion{O}{1}] 63 $\mu$m
brightness has pre-shock density $n_0=10^3$ cm$^{-3}$. Such a
pre-shock density is reasonable considering that we are observing
shocks into molecular clouds. This shock corresponds to what we called
`region M' (\S\ref{sec:cartoon}), although we suggested a 
somewhat lower pre-shock density of $10^2$ cm$^{-3}$ for this shock.
In fact we do not directly measure $n_0$ and instead estimated it
from the density in the emitting region, so a higher $n_0$ could be consistent
with the observations. The compression by the shock depends on many
unknown factors, including the strength of the preshock magnetic field.

Before stepping through the comparison of more observed results
to the models, we should clarify that the models are themselves
significantly different. The physics of shock fronts, especially
into dense gas and dust clouds, with so many modes of energy
release and interaction, is exceedingly complex and cannot be
expected to be accurately predicted by a simple theoretical model.
Let us compare the HM89 and HRH87 models for $n_0=10^3$ cm$^{-3}$ and
$V_s=100$ km~s$^{-1}$. The HRH87 model is called D100 in their
paper. The magnetic field is negligible in the D100 model, but
comparing the E100 and B100 models it appears that a stronger magnetic
field makes relatively little difference in predicted line
brightness within the context of the HRH87 model.
Comparing the line brightnesses between the HM89 and HRH87 models,
it is clear that there are very significant differences. 
An extreme case is the [\ion{O}{1}] 6300 \AA\ line, which is very
bright in the HM89 model but weak in the HRH87 model; the difference
is a factor of about 8. 
A significant part of the difference could be neglect of the higher ionization states
in the HM89 model; we suspect the [\ion{O}{2}] and [\ion{O}{3}] optical lines
are much brighter than [\ion{O}{1}].
Other significant differences are
the importance of Ly$\alpha$, many high-excitation ultraviolet lines, 
and the 2-photon continuum; all of these are very bright in the
HRH87 model. Part of the difference could be neglect of the radiative
transfer through dust in the HRH87 model; we expect most of the ultraviolet photons
are absorbed by the cooler layers of the post-shock gas and dust or by the remaining
unshocked gas and dust.

Comparing our observations to the theoretical shock models,
we reach the following conclusions:

\noindent{\it Dust emission---}We observe a far-infrared continuum 
toward the shock fronts, in excess of what we would expect from
dust in the quiescent pre-shock gas. The dust has a 
color temperature distinctly warmer than dust in molecular or atomic clouds.
This could be due to significant collisional or radiative heating in
the post-shock gas raising the equilibrium temperature of the grains,
or it could be due to a significant difference in the size distribution
of the post-shock grains. Indeed the post-shock size distribution is
predicted to be significantly different because the larger grains are
shattered (\cite{jones96}). 
The HM89 model predicts too little [\ion{Fe}{2}] 26 $\mu$m
and [\ion{Si}{2}] 35 $\mu$m emission by about a factor of 10.
This is probably due to treatment of dust destruction:
the elemental abundances in the HM89 model are set at depleted cosmic abundances
typical of what may occur in the pre-shock gas, while 
dust destruction is now predicted to be significant at these shock velocities
(\cite{jones96}). 
The HM89 model predicts some infrared continuum, but its flux is
weaker than the [\ion{O}{1}] 63 $\mu$m line, which does not agree
with our observations: we find that the continuum flux is much greater than 
that of any of the emission lines. The HRH87 is on the opposite extreme: they
use the entire cosmic abundance in the gas phase, implicitly assuming
complete destruction of the dust and predicting no infrared continuum.

\noindent{\it Infrared versus ultraviolet---}The infrared emission
escapes from the shocks and arrives at our telescopes, while the
ultraviolet (and optical) emission either never escapes the
shock or is absorbed before reaching us. The HRH87 model
predicts copious ultraviolet emission, with the strongest 
emission (in order of brightness) arising from Ly$\alpha$,
2-photon continuum, \ion{C}{3} 977~\AA, and \ion{He}{2} 304~\AA.
The sum of the 5 infrared lines they report 
is still 20 times weaker than Ly$\alpha$ and 3 times weaker than
just the \ion{He}{2} 304~\AA\ line. Thus ultraviolet lines 
(and 2-photon continuum) are by far the dominant cooling behind
radiative shocks according to the HRH87 model. On the other hand,
the HM89 model predicts a significantly different spectrum,
with relatively brighter optical and infrared lines. The brightest
line in the HM89 model is the optical [\ion{O}{1}] 6300~\AA\ line,
followed by H$\alpha$ and comparable brightnesses of 
infrared [\ion{O}{1}] 63 $\mu$m, 
optical [\ion{S}{2}] 6731~\AA\ and [\ion{N}{2}] 6560~\AA, and
ultraviolet \ion{C}{2}] 2326~\AA\ lines.
Remarkably, the brightness of the \ion{C}{2}] 2326~\AA\ and 
[\ion{O}{2}] 3726+3729~\AA\ lines,
and H$\alpha$ and H$\beta$,
agree for the two models, but there are many brighter 
ultraviolet lines in the HRH87 model. The difference between the
two models is at least partially due to the different assumptions
about dust grains. The HRH87 model has no dust grains, so that
ultraviolet transitions that are optically thick can scatter
from ion to ion and eventually escape the shock or have a
quantum decay into two photons.
If dust grains were included, the ultraviolet photons would not last long
before being absorbed. The ultraviolet photons are produced in a 
thin layer behind the shock, and their energy would heat the surrounding
gas, as a radiative precursor into the pre-shock gas and as an
enhanced radiation field in the cooler layers of the post-shock gas.
In the HM89 model, the radiative transfer does take into
account absorption of the ultraviolet lines by dust grains (\cite{mh79}),
which explains why this model predicts much weaker ultraviolet emission
than the HRH87 model. Our detection of a far-infrared continuum
suggests that the models must include the effects of dust on
the ultraviolet photons.

\noindent{\it Molecules---}We detected both the S(3) and S(9)
lines of H$_2$ (\S\ref{section:h2}),
as well as OH and weak H$_2$O emission (\cite{rr99}).
The model of HRH87 does not include molecules, so we will not discuss it here.
The model of HM89 includes a chemical reaction network, and they predict the
brightness of the S(3) and S(9) lines to be $9\times 10^{-6}$ and
$8\times 10^{-5}$ erg~s$^{-1}$~cm$^{-2}$~sr$^{-1}$, respectively.
The predicted brightness of the S(9) line is 
about half of what we observed, suggesting that a significant fraction
of the S(9) line could be produced by a J-type shock.
However, the predicted brightness of the S(3) line is far lower
than we observed. Indeed, for all 4 of the molecular shocks we
observed, the S(3) line is brighter than the S(9) line; this is
contrary to the model, which predicts S(3) about 10 times weaker
than S(9). Instead, it appears that a denser, non-dissociating shock is required
to explain the S(9) line brightness. This picture agrees with 
presence of a preshock cloud with a range of densities,
as shown in our cartoon model.
Therefore, it appears that the shock models cannot
simultaneously predict both the ionic lines and (all of) 
the molecular lines with $n_0=10^3$~cm$^{-3}$ and $V_s=100$ km~s$^{-1}$.
Instead, it appears that a denser, non-dissociating shock is required
to explain the S(3) line brightness. 

To explain the brightness and
excitation of the OH and H$_2$O emission, and the $\sim 30$ km~s$^{-1}$
width of the CO and CS lines, we suggested C-type shocks into
gas with density $> 10^4$ cm$^{-3}$ (Reach \& Rho 98, 99).
Models for such shocks were developed by Draine, Roberge, \&
Dalgarno (1983; hereafter DRD). For $n_0=10^4$ cm$^{-3}$ and
$V_s=30$ km~s$^{-1}$, the DRD models predict that H$_2$
is the dominant coolant, with the brightest lines being S(3) and S(5).
In the DRD model, the S(9) line is about 10 times weaker than the
S(3) line. 
Given that we observe S(3) and S(9) lines of comparable brightness,
it appears that we are required to have the bulk of the 
the S(3) line arise from a C-type shock and part of the 
S(9) line arise from a J-type shock. The observed surface brightness
of the S(3) line is lower than the predicted brightness from
the DRD model ($I_{pred}=3\times 10^{-3}$), but that could be
due to beam dilution of small, dense clumps that produce the C-type shock.

\noindent{\it Comparison to a model of W~44---}
Two recent papers by Cox et al. (1999) and
Shelton et al. (1999) presented a
model for the remnant W~44. 
In this model, all observed properties of the remnant, from radio
emission to $\gamma$-rays, and specifically including the infrared lines,
are produced by a blast wave into a smooth medium with a density of 
6~cm$^{-3}$. This model is in marked contrast to the one we presented
here, with pre-shock densities ranging from $< 1$ to $>10^2$ cm$^{-3}$,
and some discussion may help to resolve confusion about how they 
could both be explaining the same data. 
First, we note that the evidence for molecular cloud interaction is
very strong. Shocked CO line with widths of $\sim 30$ km~s$^{-1}$
were revealed by Seta et al.
(1998) and ourselves (this paper, Fig.~\ref{fpw44}).
\def\extra{
The older observations from the 1970's 
(\cite{wootten}) were inconclusive to some researchers (\cite{denoyer}),
but that was largely because of the complicated mix of optically-thick
pre-shock gas with shocked gas along the line of sight (cf. discussion 
in \cite{rr99}). The more recent observations clearly reveal the interaction
between the supernova remnant and the molecular cloud.
}
Second, we note that the uniform pre-shock conditions in the Cox et al. model are 
not realistic in a large region of the multiple-phase interstellar medium.

In our model, we attribute the molecular line observations of W~44,
including H$_2$ near-infrared emission (\cite{seta}), infrared CO lines (\cite{rr98}),
wide millimeter CO and other lines (\cite{seta}), and 
radio OH 1720 MHz masers (\cite{claussen}),
to dense, shocked gas.
These lines cannot be produced by an ionized low-density post-shock region. 
The analogy to IC~443 is useful---W~44 is very similar in many regards, 
with similar brightnesses of the 63 $\mu$m line and brightnesses and
widths of the millimeter-wave CO (and other) lines---and few would
deny that IC~443 is interacting with a molecular cloud.
Finally, the observed brightness of the [\ion{O}{1}] 63 $\mu$m line
cannot easily be explained by a low-density medium, in contrast to
the discussion claimed by Cox et al. (1999). 
The excitation of [\ion{O}{1}] evidenced by the 145/63 $\mu$m
ratio requires a higher density region (Fig.~\ref{fig:excite}). 
Obviously, [\ion{O}{1}] also requires
a neutral region, for if the H is ionized the O will also be ionized.
The observed lines were so `shockingly' bright that we concluded immediately
that they were due to a dense shock (\cite{rr96}). In this paper we considered
a wide range of densities and ruled out the low densities because
(1) they are inconsistent with the line ratios, and (2) the path length
required through a low-density gas would be longer than the remnant size
in order to build up enough column density to produce the observed line
brightness. Cox et al. reduced the brightness of the 63 $\mu$m by a factor
of 100 to match their models. One of the reduction factors is a
a factor of 10 to match the estimated average over the entire remnant. 
Another factor applied by Cox et al. was to enhance their
prediction because of limb brightening and local rippling of the shell.
While this is perfectly reasonable, this correction was applied to
the low-resolution 63 $\mu$m line observations, which trace relatively
cool gas, and not to the higher-resolution H$\alpha$ observations 
(\cite{giacani}), which trace warmer gas that might be even
more edge brightened.
All of these factors probably cancel out to first order when the average 
over our large beam is made, so that the [\ion{O}{1}]/H$\alpha$ ratio is
not matched by their model.

The truth is probably that there
were both high and low density regions in the interstellar medium around
the W~44 progenitor star (as considered in this paper and by \cite{chevalier}). 
The particular lines of sight studied in 
this paper are centered on a special regions,
where evidence for a dense shocks are overwhelming. Away from these positions,
it is likely that the lower-density shocks contribute some of the
low-level 63 $\mu$m emission. 
In Figure~\ref{fig:cartoon}, we are pointed right at a region C clump, while
the average over a large region would include different regions in different
proportions. Because the 63 $\mu$m line brightness is expected
to scale approximately as $n_0 V_s$ (HM89), the denser regions 
contribute relatively more to the observed brightness, when they are present.
Further observational work may be able to separate the dense shocked gas
from the more rarefied and widespread interclump gas, using imaging to
resolve the clumpy spatial distribution and spectroscopy to
separate the narrower line widths.

\section{Conclusions}

Based on {\it ISO} spectroscopic observations
of supernova remnants interacting with molecular clouds, we found
evidence for a wide range of pre- and post-shock conditions; these
properties are summarized in Table~\ref{tab:cartoon} and illustrated in 
Figure~\ref{fig:cartoon}. 
Of the three density regimes that we identified, most of the infrared
emission arose from the shocks into moderate-density molecular gas
(`region M'). The energy of these shocks is radiated via continuum
emission from surviving dust grains and [\ion{O}{1}] 63 $\mu$m and
[\ion{Si}{2}] and other infrared atomic fine structure lines. 
The dust continuum was difficult to separate from the far-infrared
emission from cold, unrelated dust, but the shocked dust was evident at 
80--100 $\mu$m as a warmer component of the spectrum. Some of the dust
grains must have been destroyed, because we observe bright emission
lines from dust `vapors,' including the [\ion{Fe}{2}] 26 $\mu$m line.
To produce the observed line dust vapor lines, we require that 
1/3 of the Si and Fe-bearing dust mass was vaporized in the shocks.
Using the Fabry-Perot observations, the width of the 63 $\mu$m line
was found to be $\sim 100$ km~s$^{-1}$. Theoretical models can reproduce
the brightness of the 63 $\mu$m line with such shock velocities and
the properties of `region M' (\cite{HM89}); theoretical models can also
reproduce the observed amount of grain destruction (\cite{jones96}).
Higher-density gas was required to explain the bright H$_2$ line emission,
leading to our `region C,' for clumps. Lower-density gas was required to 
explain the higher-ionization lines, leading to our `region A,' for atomic
gas. 

A general conclusion from these observations is that molecular shock
fronts are copious producers of infrared emission. We anticipate many
applications of these types of observations for further understanding
the nature of molecular supernova remnants, such as the fate of dense
clumps, the fate of dust grains, and the effect of cooling on the 
remnant evolution. In turn, these studies can
elucidate, to some extent, the nature of the molecular clouds before the 
shocks. If molecular clouds were uniform, or nearly so, then we would
expect an orderly progression of pre-shock to post-shock gas, with
radiative coolants characteristic of a single shock velocity. On the
contrary, we observe emission from coolants as diverse as multi-atom
molecules, multiply-ionized atoms, and dust grains and their vapors.

\acknowledgements 
We would like to thank Pierre-Olivier Lagage for helpful discussions in
interpreting atomic fine structure lines, and we thank Emmanuel Caux,
Steve Lord, and Sergio Molinari for clarifying details of the LWS 
instrument and calibration. 


\begin{thebibliography}{} 
 
\small

\bibitem[Anders \& Grevesse 1989]{anders} Anders, E., \& Grevesse, N. 1989,
	Geochim. Cosmo. Acta, 53, 197
\bibitem[Arendt 1996]{arendtcatalog} Arendt, R. G. 1996, ApJS, 70, 181
\bibitem[Arikawa et al. 1999]{arikawa} Arikawa, Y., Tatematsu, K., Sekimoto, Y.,
\& Takahashi, T., PASJ, 51, L7
\bibitem[Becklin 1997]{becklin} Becklin, E. E. 1997, in {\it The 
	Far-infrared and Submillimeter Universe}, ESA SP-401, p. 201
\bibitem[Blum \& Pradhan 1992]{blum} Blum, R. D., \& Pradhan, A. K. 1992, ApJS 80, 425
\bibitem[Cesarsky et al. 1999]{cesarsky443} Cesarsky, D., Cox, P., et al. 1999, preprint
\bibitem[Chevalier 1999]{chevalier} Chevalier, R. A. 1999, ApJ, 511, 798
\bibitem[Claussen \etal 1997]{claussen} Claussen, M. J., Frail, D. A., Goss,
 W. M., \& Gaume, R. A. 1997, ApJ, 489, 143
\bibitem[Clegg et al. 1996]{clegg} Clegg, P. E., et al. 1996, A\& A, 315, L38
\bibitem[Cox et al. 1999]{cox99} Cox, D. P., Shelton, R. L., Maciejewski, W.,
Smith, R. K., Plewa, T. , Pawl, A., \& R\'ozyczka, M. 1999, ApJ, 524, 179
\bibitem[de Graauw et al. 1996]{swsref} de Graauw, T., et al. 1996,
{\it Astron. Astrophys.} {\bf 315}, L49--L54.
\bibitem[DeNoyer 1983]{denoyer} DeNoyer, L. K. 1983, ApJ, 264, 141
\bibitem[D\'esert, Boulanger, \& Puget 1990]{dbp} D\'esert, F.-X., Boulanger, F., \&
Puget, J.-L. 1990, A\& A, 237, 1
\bibitem[Draine \& Lee 1984]{drainelee} Draine, B. T., \& Lee, H. M. 1985, ApJ, 285, 89
\bibitem[Draine et al. 1983]{DRD} Draine, B. T., Roberge, W. G., \& Dalgarno, A. 1983, ApJ, 264, 485 
\bibitem[Fitzpatrick 1996]{fitzpatrick} Fitzpatrick, E. L. 1996, \apjl, 473, L55 
\bibitem[Frail et al. 1994]{Frail94} Frail, D. A., Goss, W. M., Slysh, V. I. 1994, ApJ, 424, L111
\bibitem[Frail et al. 1996]{Frail96}  Frail, D. A., Goss, W. M., Reynoso, E. M., Giacani, E. B., Green, A. J., 
	Otrupcek, R. 1996, AJ, 111, 1651 
\bibitem[Froese Fischer 1983]{froese} Froese Fischer, C. 1983, J. Phys. B, 16, 157
\bibitem[Giacani et al. 1997]{giacani} Giacani, E. B., Dubner, G. M., Kassim, N. E., 
Frail, D. A., Goss, W. M., Winkler, P. F., \& Williams, B. F. 1997, AJ, 113, 1379
\bibitem[Hollenbach \& McKee 1979]{mh79} Hollenbach, D. J., and McKee, C. F. 1979, ApJS, 41, 555
\bibitem[Hollenbach \& McKee 1989]{HM89} Hollenbach, D. J., and McKee, C. F. 1989, ApJ, 342, 306
\bibitem[Hollenbach, Werner, \& Salpeter 1971]{h2form} Hollenbach, D. J., Werner, M. W.,
\& Salpeter, E. E. 1971, ApJ, 163, 165
\bibitem[Jones, Tielens, \& Hollenbach 1996]{jones96} Jones, A. P., Tielens, 
A. G. G. M., \& Hollenbach, D. J. 1996, ApJ, 469, 740
\bibitem[Kessler et al. 1996]{kessler} Kessler, M. F. et al. 1996, A\& A, 315, L27
\bibitem[Koo \& Heiles 1995]{kooheiles} Koo, B.-C., \& Heiles, C. 1995, ApJ, 442, 679
\bibitem[Krueger \& Zsyzak 1970]{piicoll} Krueger, T.K., and Czyzak, S.J. 1970,
	Proc. Roy. Soc. London A, 318, 531 
\bibitem[Lagache et al. 1998]{lagachecold} Lagache, G., Abergel, A., Boulanger, F.,
	\& Puget, J.-L. 1998, A\& A, 333, 709
\bibitem[Launay \& Roueff 1977]{launay} Launay, J. M. \& Roueff, E. 1977, A\& A, 56, 289
\bibitem[Liu et al. 1996]{liu} Liu, X.-W. et al. 1996, A\&A, 315, L257
\bibitem[Lockett et al. 1998]{lockett} Lockett, P., Gauthier, E., \& Elitzur, M.
1999, ApJ, 511, 235
\bibitem[Mendoza \& Zeippen 1982]{mendoza} Mendoza, C., \& Zeippen, C. J. 1982,
	MNRAS, 199, 1025
\bibitem[Nussbaumer \& Storey 1982]{nussbaumer} Nussbaumer, H., \& Storey, P.J. 1982, A\& A, 113, 21
\bibitem[Oliva et al. 1999a]{oliva443} Oliva, E., Lutz, D., Drapatz, S.,
	Moorwood, A. F. M. 1999a, A\& A, 341, L75
\bibitem[Oliva et al. 1999b]{oliva} Oliva, E., Moorwood, A. F. M, Drapatz, S.,
	Lutz, D., \& Sturm, E. 1999b, A\& A, 343, 943
\bibitem[Pickett et al. 1996]{pickett} Pickett, H. M. Poynter, R. L., Cohen, E. A., Delitsky, M. L., 
	Pearson, J. C., \& M\"uller, H. S. P. 1996, {\it Submillimeter, Millimeter, and
	Microwave Spectral Line Catalog: Revision 4}, JPL publication 80-23,
	(Pasadena: JPL)
\bibitem[Poglitsch 1998]{firstref} Poglitsch, A. 1998, in {\it The Far-Infrared and
	Submillimeter Universe} (ESA SP-401), 
	eds. G. Pilbratt, S. Volonte, \& A. Wilson (ESA: Noordwijk), in press
\bibitem[Reach \& Boulanger 1998]{reachboul} Reach, W. T., Boulanger, F. 1998, in
{\it The Local Bubble and Beyond}, eds. D. Brietschwerdt, M. J. Freyburg, \&
J. Tr\"umper (Berlin: Springer), 353
\bibitem[Reach \& Rho 1996]{rr96} Reach, W. T., \& Rho, J.-H. 1996, A\& A, 315, L277 (paper I)
\bibitem[Reach \& Rho 1998]{rr98} Reach, W. T., \& Rho, J.-H. 1998, ApJ, 507, L93
\bibitem[Reach \& Rho 1999]{rr99} Reach, W. T., \& Rho, J.-H. 1999, ApJ, 511, 836
\bibitem[Reach et al. 1995]{reachfiras} Reach, W. T., et al. 1995, ApJ, 451, 188
\bibitem[Rho 1995]{rhothesis} Rho, J.-H. 1995, 
An X-Ray Study of Composite Supernova Remnants (Ph.D. thesis, Univ. of Maryland)
\bibitem[Rho \& Petre 1998]{RP98} Rho, J.-H., \& Petre, R., 1998, \apjl, 503, L167
\bibitem[Rho et al. 2000]{rho00} Rho, J., Jarrett, T., Cutri, R., Reach, W., \&
	van Dyk, S. 2000, submitted to ApJ
\bibitem[Richter, Graham, \& Wright 1995]{richter} Richter, M., Graham, J. R.,
	\& Wright, G. S. 1995, ApJ, 454, 277
\bibitem[Savage \& Sembach 1996]{sembachsavage} Savage, B. D., \& Sembach, K. R. 1996,
ARAA, 34, 279
\bibitem[Schmutzler \& Tscharnuter 1993]{nei} Schmutzler, T. \& Tscharnuter, W. M. 1993, \aap, 273, 318 
\bibitem[Seta et al. 1998]{seta} Seta, M. et al. 1998, ApJ, 505, 286
\bibitem[Shelton et al. 1999]{shelton} Shelton, R. L., Cox, D. 
P., Maciejewski, W. , Smith, R. K., Plewa, T. , Pawl, A.  \& R\'ozyczka, M.  
1999, \apj, 524, 192 
\bibitem[Spitzer 1978]{spitzer} Spitzer, L. 1978, {\it Physical Porcesses in the 
Interstellar Medium} (New York: Wiley)
\bibitem[Wardle 1999]{wardle} Wardle, M. 1999, \apjl, 525, L101
\bibitem[Wild 1995]{wild} Wild, W. 1995, {\it The 30m Manual: A Handbook for the IRAM 30m
	Telescope, Pico Veleta, Spain}
\bibitem[Wootten 1977]{wootten} Wootten, A. 1977, ApJ, 216, 440
\end{thebibliography}
\end{document}